\documentclass[12pt]{article}

\usepackage[dvips]{graphicx}
\usepackage{amsmath}
\usepackage{amsfonts}
\usepackage{amssymb}

\newcommand{\be}{\begin{equation}}
\newcommand{\ee}{\end{equation}}
\newcommand{\bea}{\begin{eqnarray}}
\newcommand{\eea}{\end{eqnarray}}

\newcommand{\nn}{\nonumber}

\begin{document}

\begin{flushright}
\small
DESY 12-073 \\
CERN-PH-TH/2012-127\\
\end{flushright}

\vskip 8pt

\begin{center}
{\bf \LARGE {
On the gauge dependence of vacuum\\[0.6ex] transitions at finite temperature
 }}
\end{center}

\vskip 12pt

\begin{center}
\small
{\bf Mathias Garny$^{a}$, Thomas Konstandin$^{a,b}$ } \\
--\\
{\em $^a$ DESY, Notkestr.~85, 22607 Hamburg, Germany } \\
{\em $^b$ CERN Theory Division, 1211 Geneva, Switzerland } \\
\end{center}

\vskip 20pt

\begin{abstract}
\vskip 3pt
\noindent

In principle, observables as for example the sphaleron rate or the
tunneling rate in a first-order phase transition are
gauge-independent. However, in practice a gauge dependence is
introduced in explicit perturbative calculations due to the breakdown of the
gradient expansion of the effective action in the symmetric phase.  We
exemplify the situation using the effective potential of the Abelian
Higgs model in the general renormalizable gauge. Still, we find that the
quantitative dependence on the gauge choice is small for gauges that
are consistent with the perturbative expansion.

\end{abstract}

\newpage

\section{Introduction\label{sec:intro}}

Functional methods are an indispensable tool in studying the physics
of spontaneous symmetry breaking and phase
transitions~\cite{Coleman:1973jx, Dolan:1973qd, Jackiw:1974cv,
  Dolan:1974gu}. The main advantage in this approach is that the
effective action encodes the ground state of the system in a
transparent way. On the technical side, it facilitates the resummation
of tadpole diagrams and reduces perturbative calculation to the subset
of one-particle-irreducible diagrams.

A large drawback of the effective action is however that it is not
explicitly gauge-independent~\cite{Nielsen:1975fs, Fukuda:1975di,
  Aitchison:1983ns, Kobes:1990dc, Buchmuller:1994vy,
  Buchmuller:1995sf}. This makes it necessary to distinguish between
the gauge dependence that is expected from the one that is introduced
by approximation schemes that break the gauge invariance
additionally. The first class of gauge dependences is well represented
by the Nielsen identities while the second can for example arise from
the use of a loop expansion in perturbation theory or from the
expansion of the effective action in gradients.

The aim of the present work is to disentangle these different sources
of gauge dependence for vacuum transitions~\cite{Metaxas:1995ab} at
finite temperature in the Abelian Higgs model. After a general
introduction to the model (Section \ref{sec:model}) and the Nielsen
identities (Sections \ref{sec:nielsen} and \ref{sec:gradexp}), the
effective potential is calculated (Section \ref{sec:effpot}). The main
focus is hereby on resummation of infrared effects at finite
temperature. In this section, it is also explicitly demonstrated that
the position of the minimum of the effective potential transforms
under changes of the gauge fixing parameter $\xi$ according to the Nielsen
identity. Subsequently (Section \ref{sec:grad_check}), the same is
demonstrated for the (off-shell) effective action in the gradient
expansion. Finally (Sections \ref{sec:tunnel} and
\ref{sec:sphaleron}), the gauge dependence of the tunneling action and
the sphaleron energy are discussed. In these cases an additional
complication arises, namely the breakdown of the gradient expansion in
the symmetric phase. We discuss, to what extend the gauge dependence
of these phenomena can be quantified and potentially reduced before we
conclude (Section \ref{sec:conclusion}).

\section{The model\label{sec:model}}

In the following we lay out the details of the model under
consideration. In order to simplify the analysis and to focus on the
main impact of the gauge dependence of the effective action, we
discuss an Abelian Higgs model. This model has all necessary
ingredients that occur in the Standard Model but does not contain
fermions or the correct symmetry breaking pattern as observed in
Nature. Nevertheless, for the order of perturbation theory we work at
these features are not important and the presented arguments can
immediately be carried over to the Standard Model case.

The Lagrangian is given by
\be
{\cal L} = D_\mu \Phi^* D^\mu \Phi - \frac14 F_{\mu\nu} F^{\mu\nu} - V(\Phi^* \Phi) \, ,
\ee
with the scalar potential
\be
V(\Phi^* \Phi) = - \mu^2 \Phi^* \Phi + \frac{\lambda}{4} (\Phi^* \Phi)^2
= \frac{\lambda}{4} (\Phi^* \Phi - v^2)^2 .
\ee
The gauge covariant derivative is given by $D_\mu \Phi
= (\partial_\mu - i g A_\mu) \Phi $, while the field strength
tensor reads $F_{\mu\nu} = \partial_\mu A_\nu - \partial_\nu A_\mu$.

We are interested in the effective action $\Gamma$ that is
obtained~\cite{Coleman:1973jx, Jackiw:1974cv} by
Legendre transformation of the generating functional of connected
Greens functions $W(j)$
\bea
\label{eq:defGamma}
e^{i W(j)} &=& \int {\cal D} \Phi {\cal D} A \,
e^{i \int d^4 x {\cal L}(\Phi) + j \, \Phi} \, ,\nn \\
\Gamma(\phi) &=& \left. W(j) - \int d^4x \, j \, \phi \right|_{\phi = dW/dj}.
\eea
This implies after a shift in the integration variable $\Phi \to \Phi + \phi$
\be
e^{i \Gamma} = \int {\cal D} \Phi {\cal D} A \,
e^{i \int d^4 x {\cal L}(\Phi+\phi) + j \, \Phi} \, .
\ee
By construction the field in the shifted theory does not have a vacuum
expectation value (vev). Hence, the bi-linear term containing the
source $j$ has to cancels all tadpole diagrams and beyond this
cancellation, the source $j$ has no impact on the effective action.
Therefore $\Gamma$ contains only the connected
one-particle-irreducible vacuum diagrams in the shifted theory. The
effective potential is obtained by restricting the effective action
to homogeneous field expectation values $\phi$,
$V_{eff} = \Gamma/\int d^4x$.

For perturbative calculations, the Lagrange density has to be
supplemented by a gauge fixing procedure that results in an additional
gauge fixing term and a contribution from the Fadeev-Popov ghosts.

A common choice is to use the general Lorentz gauge 
\bea
\textrm{Lorentz gauge}: \quad
{\cal L}_{gf} &=& - \frac{1}{2\xi} (\partial_\mu A^\mu)^2 \, , \quad \nn \\
{\cal L}_{FP} &=& - \bar c \, \square \, c \, .
\eea
In this gauge the Fadeev-Popov ghosts do not couple to the remaining
particles, but the Goldstone boson and the unphysical $k^\mu$
polarization of the gauge boson mix after spontaneous symmetry
breaking.  Nevertheless, this problem is removed in the Landau gauge,
$\xi \to 0$.

Another common choice is the $R_\xi$-gauge \cite{Fujikawa:1972fe}
\bea
R_\xi\textrm{-gauge}: \quad
{\cal L}_{gf} &=& - \frac{1}{2\xi} 
(\partial_\mu A^\mu+ ig \xi (\phi^* \Phi - \Phi^* \phi) )^2 \, , \quad \nn \\
{\cal L}_{FP} &=& -\bar c \, ( \square + \xi g^2 (\phi^* \Phi + \Phi^* \phi))  \, c \, .
\eea
The contribution containing the background field $\phi$ is designed
to cancel the mixing term between the Goldstone mode and the $k^\mu$
polarization of the gauge boson thus partially diagonalizing the
propagator structure of the theory. However, having an explicit
dependence on the background field in the gauge fixing term leads to
some complications. 

First of all, in Lorentz-gauge a shift in $\phi$ is equivalent to
attaching an external $\Phi$ field to the
diagram~\cite{Weinberg:1973ua, Dolan:1974gu} (this follows immediately
from inspecting the Feynman rules). Hence the tadpole can be obtained
from the derivative of the effective action.  In general gauges this
only holds in the minimum of the potential (where the tadpole has to
vanish) while in Lorentz gauge this is also true away from the extrema
of the potential.

One consequence of this fact is that the two-point functions of the
Higgs and Goldstone bosons (at vanishing external momentum) can be
obtained from the second derivatives of the effective potential. In
particular, the Goldstone bosons are massless in the broken phase
(what can be a curse or a blessing). In $R_\xi$-gauge the Goldstone bosons
have a mass already on tree-level for $\xi \not= 0$.

Furthermore, any tadpole that is generated in a higher loop
perturbative calculation can in Lorentz gauge be absorbed by adjusting
the appropriate counterterms. This is particularly handy in zero
temperature calculations where the Higgs vev is an input observable
parameter and kept fixed order by order in the perturbative
expansion. Especially when several background fields are present, it
is not guaranteed in the $R_\xi$-gauge that the tadpoles can be
absorbed into the counterterms of the Lagrangian what makes it
technically more difficult to keep the vevs fixed in higher loop
calculations \cite{Weinberg:1973ua}. Note that at finite temperature
this feature is not so important, since part of the shift of the vev
by loop effects is physical and one has to deal with a
temperature-dependent vev anyway.

The behavior of the effective action under gauge transformations was
originally analyzed in the Lorentz gauge by
Nielsen~\cite{Nielsen:1975fs}.  Similar relations have been derived
shortly after for $R_\xi$-gauges \cite{Fukuda:1975di,
  Aitchison:1983ns, Kobes:1990dc} but their application is somewhat
more involved.  Still the general picture is the same as in the case
of the original Nielsen identities: While the value of the effective
action in its extrema is gauge-independent, the position of the
extrema can have a gauge dependence. This implies that the critical
temperature of the phase transition is gauge-independent. For vacuum
transitions, the configurations that extremize the effective action
can be gauge-dependent while the value of the action for the
configuration is not. The prove of this is sketched in the next
section.

Ultimately, we aim to discuss the gauge dependence of the sphaleron
energy and the tunneling rate.  In this work we focus on the
$R_\xi$-gauge for two reasons. First, we find it technically easier to
eliminate the mixing between the gauge boson and the Goldstone boson and to
deal with the peculiarities of the $R_\xi$-gauge than to perform loop
calculations in the general Lorentz gauge
\footnote{See \cite{Metaxas:1995ab} for a zero temperature analysis of
  the gauge dependence of the effective action in general Lorentz
  gauge.}.  Second, the convergence of the perturbation theory is in
the Lorentz gauge less obvious than in the
$R_\xi$-gauge~\cite{Laine:1994bf, Laine:1994zq, Kripfganz:1995jx}.

\section{Nielsen identity in $R_\xi$ gauge
  \label{sec:nielsen}}

In this section we briefly review the Nielsen identities for the
effective potential, paying special attention to the additional
complications arising in the $R_\xi$-gauge. In order to derive the
Nielsen identity for a general gauge fixing we write the latter in the
form,
\bea
{\cal L}_{gf} &=& -\frac{1}{2\xi}F^2 \,,\nn \\
{\cal L}_{FP} &=& -\bar c \, \left( \frac{\delta F}{\delta A_\mu}
\partial_\mu + \frac{\delta F}{\delta \Phi}ig\Phi 
+ \frac{\delta F}{\delta \Phi^*}(-ig\Phi^*) \right) \, c \,,
\eea
where $F=\partial_\mu A^\mu$ in Lorentz gauge and $F=\partial_\mu
A^\mu+ ig \xi (\tilde\phi^* \Phi - \Phi^* \tilde\phi)$ for the
$R_\xi$-gauge. For the derivation of the Nielsen identities, it is
convenient to discriminate the field expectation value
$\phi\equiv\langle\Phi\rangle$ that appears due to the Legendre
transformation of the generating functional from the explicit
dependence on the background field in the gauge fixing term, which is
denoted by $\tilde\phi$ here and can be considered as an external
parameter at this stage of the calculation. Both fields will be
identified in the end, but for now we distinguish them.

The dependence on $F$ in general leads to a gauge dependence of the
effective action $\Gamma$. For an arbitrary change in the gauge fixing
term parametrized by $F\to F+\delta F$ and $\xi\to \xi+\delta\xi$ the
change of the functional $W$ is given by~\cite{Kobes:1990dc}
\be\label{eq:DelW}
\delta W = {} - i\int d^4x\int d^4y \, j_i \, 
\langle \delta_g\Phi_i(x) c(x)\bar c(y) \delta' F(y) \rangle \, ,
\ee
where $\delta' F\equiv \delta F - F/(2\xi)\delta\xi$. Furthermore,
we collectively denote by $\phi_i$ all field expectation values that the
effective action depends on, and by $\delta_g\Phi_i$ the variation of
the corresponding field operator under a gauge transformation
\be
\delta_g\Phi = ig\Phi,\  \delta_g\Phi^* = -ig\Phi^*,\  \delta_g A_\mu =
\partial_\mu \;.
\ee
The expectation value of an operator $\langle \cal{O } \rangle $ is
defined in the usual path integral sense.

The functional $W$ is gauge invariant up to the source term involving
$j_i$ and the gauge fixing term ${\cal L}_{gf} + {\cal L}_{FP}$. The
identity (\ref{eq:DelW}) reflects the fact that a change in the gauge
fixing functional can via a gauge transformation shifted into a change
of the source term. Now, notice that for any external parameter
$\lambda$ one has the relation
\be
\left. \frac{d\Gamma (\phi, \lambda)}{d\lambda} \right|_{\phi = const}
= \left. \frac{d W (j, \lambda)}{d\lambda} \right|_{j = const} \, ,
\ee
what can be verified using the definition (\ref{eq:defGamma}). The
gauge fixing parameters $\xi$ and $\tilde \phi$ are external
parameters such that one can translate (\ref{eq:DelW}) into a relation
for $\Gamma$ if the left hand side is understood as being varied with
$\phi$ kept fixed, namely
\be\label{eq:DelGamma}
\delta\Gamma = i\int d^4x\int d^4y \frac{\delta\Gamma}{\delta\phi_i(x)}
\langle \delta_g\Phi_i(x) c(x)\bar c(y) \delta' F(y) \rangle \, .
\ee
This immediately yields the Nielsen identity for the effective action,
which results when only considering a change in the gauge parameter $\xi$,
\be\label{eq:Nielsen}
\xi\frac{\partial\Gamma}{\partial\xi} = -\int d^4x \left\{ 
\frac{\delta\Gamma}{\delta A_\mu(x)} C_{A^\mu}(x) 
+ \frac{\delta\Gamma}{\delta \phi(x)} C_\phi(x) 
+ \frac{\delta\Gamma}{\delta \phi^*(x)} C_{\phi^*}(x) \right\} \,,
\ee
where the coefficients, taking a possible $\xi$ dependence of $F$ into
account, are given by
\bea
C_{A^\mu}(x) &=& \frac{i}{2} \int d^4y \langle 
\partial_\mu c(x)\bar c(y) (F(y)-2\xi\partial F(y)/\partial\xi) \rangle \,, \nn \\
C_{\phi}(x)   &=& \frac{i}{2} \int d^4y \langle 
ig\Phi(x) c(x)\bar c(y) (F(y)-2\xi\partial F(y)/\partial\xi) \rangle \,, \nn \\
C_{\phi^*}(x)   &=& \frac{i}{2} \int d^4y \langle 
(-ig\Phi^*(x)) c(x)\bar c(y) (F(y)-2\xi\partial F(y)/\partial\xi) \rangle \,.
\eea
For a constant field expectation value, chosen to lie along the real
axis $\phi=\phi^*$, and vanishing background gauge field
$A_\mu=0$, one obtains the well-known Nielsen identity for the
effective potential,
\be
\label{eq:Nielsen_pot}
  \xi\frac{\partial V_{eff}(\phi)}{\partial\xi} + C_0 \frac{\partial
  V_{eff}(\phi)}{\partial\phi} = 0 \,. 
\ee 
The coefficient $C_0$ is obtained when evaluating
$C(x) \equiv C_{\phi}+C_{\phi^*}|_{\phi=\phi^*}$ for a constant field expectation
value $\phi$. In $R_\xi$-gauge, the general expression for $C(x)$ reads
\be
\label{eq:exp_C}
C(x) = \, \frac{ig}{2 \sqrt{2} }\, \int d^4y
 \left<
\bar c(x) \chi(x) c(y)\left(\partial_\mu A^\mu(y) 
+ \sqrt{2} g \xi \tilde\phi \chi(y) \right)
\right> \,,
\ee
while in Lorentz gauge the second summand is absent. Here we have inserted
the decomposition $\Phi=\phi+(h+i\chi)/\sqrt{2}$ where $h$ and $\chi$
denote the Higgs and Goldstone field operators, respectively.
Note that the restriction to one real background field and vanishing
gauge fields can lead to spurious minima in the effective potential
but is justified in our class of gauges \cite{Fukuda:1975di}.  This is mostly
due to the invariance of the full Lagrangian (including gauge fixing
and sources) under the transformation $\Phi\to\Phi^*$ and $A_\mu\to-A_\mu$.

\begin{figure}[t!]
\begin{center}
\includegraphics[width=0.25\textwidth, clip ]{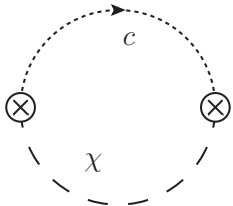}
\end{center}
\caption{
\label{fig:C}
\small One-loop diagram contributing to the Nielsen coefficient $C_0$.}
\end{figure}

For the Lorentz gauge these identities can be used directly when
inserting the gauge fixing term $F=\partial_\mu A^\mu$. In the
$R_\xi$-gauge, the effective action depends in addition parametrically
on the external field $\tilde\phi$,
i.e. $\Gamma=\Gamma[\phi;\tilde\phi]$. We are ultimately
interested in deriving a Nielsen identity for the effective action
$\bar\Gamma[\phi]\equiv\Gamma[\phi;\tilde\phi]|_{\tilde\phi=\phi}$. Compared
to the Lorentz gauge the complication arises that the field
derivatives appearing in Eq.~(\ref{eq:Nielsen}) act only on $\phi$,
while the field derivative of $\bar\Gamma[\phi]$ contains also a
part proportional to the derivative of the effective action with
respect to the background field $\tilde\phi$. Nevertheless, it is
possible to derive a Nielsen identity also for
$\bar\Gamma[\phi]$. The key observation is that the dependence on
the background field $\tilde\phi$ enters only via the gauge fixing
term \cite{Kobes:1990dc}. For that reason, the change of the effective
action when varying the background field also obeys a Nielsen
identity, which can be directly obtained from Eq.~(\ref{eq:DelGamma}),
\be
\frac{\delta\Gamma}{\delta\tilde\phi(y)} = i\int d^4x \frac{\delta\Gamma}{\delta \phi_i(x)}\left\langle \delta_g\Phi_i(x) c(x)\bar c(y) \frac{\partial F(y)}{\partial \tilde\phi} \right\rangle \,.
\ee
An analogous relation holds for the derivative with respect to the
complex conjugated field. One can use the relation above to express
the derivatives of $\bar\Gamma$ in terms of the derivatives of
$\Gamma$ with respect to the field expectation values $\phi$,
$\phi^*$ and $A_\mu$,
\be
\frac{\delta\bar\Gamma}{\delta\phi_i(y)} = \int d^4x \sum\limits_{j=A_\mu,\phi,\phi^*} 
{\cal C}_{ij}(y,x) \frac{\delta\Gamma}{\delta\phi_j(x)}\big|_{\tilde\phi=\phi} \,,
\ee
where 
\be {\cal C}_{ij}(y,x)=\delta_{ij}\delta(x-y)+i \left<
\delta_g\Phi_j(x) c(x)\bar c(y) \frac{\partial F(y)}{\partial
  \tilde\phi_i} \right> \, .  \ee
These relations imply that, if the fields fulfill the equation of
motions derived from the effective action
$\Gamma[\phi;\tilde\phi]$, i.e. $\delta\Gamma/\delta\phi_j=0$,
then the same field configurations also correspond to a stationary
point of the effective action $\bar\Gamma[\phi]$. In addition, if
${\cal C}$ is invertible, it follows that the effective action
$\bar\Gamma[\phi]$ also fulfills a Nielsen identity which is of the
same form as Eq.~(\ref{eq:Nielsen}) except that the coefficients are
replaced by the coefficients $\bar C_i$ given by
\be
 \bar C_i(y) = \int d^4x\, C_j(x)\, ({\cal C}^{-1})_{ji}(x,y)  \,,
\ee
for $i=A_\mu,\phi,\phi^*$.  Concretely, for the case of the
$R_\xi$-gauge considered here, and assuming a vanishing background
gauge field, one has
\bea
 {\cal C}_{ij}  &=&  \delta_{ij}\delta(x-y)  \\ 
&& + \quad ig^2\xi \left( \begin{array}{cc} 
 \langle \Phi(x) c(x)\bar c(y) \Phi^*(y) \rangle 
& -\langle \Phi^*(x) c(x)\bar c(y) \Phi^*(y) \rangle \\ 
-\langle \Phi(x) c(x)\bar c(y) \Phi(y) \rangle 
& \langle \Phi^*(x) c(x)\bar c(y) \Phi(y) \rangle \, .
\end{array} \right) \nn
\eea
When considering an expectation value that is constant in space-time,
the right-hand side can depend only on the difference of the
coordinates, $x-y$, and the equations can be simplified.  The
resulting Nielsen identity for the effective potential for constant
and real field expectation value in $R_\xi$-gauge has then the form
\be
\label{eq:newNielsen}
  \xi\frac{\partial \bar V_{eff}(\phi)}{\partial\xi} 
+ \bar C_0 \frac{\partial \bar V_{eff}(\phi)}{\partial\phi} = 0 \,,
\ee
where $\bar V_{eff}(\phi)=V_{eff}(\phi,\tilde\phi)|_{\tilde\phi=\phi}$, and 
\be
 \bar C_0 = C_0 \left(  1 - i g^2\xi \int d^4(x-y) 
\langle \chi (x) c(x)\bar c(y) \chi(y) \rangle \right)^{-1} \;.
\ee
The leading contribution arises at one-loop level. For a constant background
field, only a single diagram contributes to $C_0$ in the $R_\xi$-gauge, which can be
represented by the diagram shown in Fig~\ref{fig:C}. The coefficients $\bar C_0$ and
$C_0$ differ only at higher orders, which we do not consider in the following.
Therefore, we use the notation $C_0$ also in the case of
$R_\xi$-gauge.  Before checking (\ref{eq:newNielsen}) explicitly in
Section~\ref{sec:effpot}, we give attention to an analogous identity
for the kinetic term in the next section.

\section{The Gradient Expansion of the Nielsen Identity\label{sec:gradexp}}

As we have just discussed, the effective action contains a certain
gauge dependence. Nevertheless, the action evaluated in a saddle point
is gauge-independent what is for homogeneous fields reflected by the
Nielsen identities. This is explicitly demonstrated in the
subsequent section for the broken phase of the effective potential
where the convergence of perturbation theory is well under control.
However, the sphaleron energy is not a spatially constant
configuration and in particular also depends on the effective action
evaluated in the symmetric phase. This is problematic, since even
though the sphaleron energy is gauge-independent in principle, a
gauge dependence can be introduced due to the breakdown of the
gradient expansion. Still, even for configurations that do not fulfill
the equations of motion, a gauge-transformation acts on the
(off-shell) effective action according to the Nielsen identity. In the
following we derive this relation for the effective action in gradient
expansion following \cite{Metaxas:1995ab}.

To be specific, consider the effective action in gradient expansion
\be
\Gamma = \int d^4x \, \left( Z(\phi) \, \partial_\mu \phi \partial^\mu \phi  
- V_{eff}(\phi) + {\cal O}(\partial^4) \right),
\ee
where we introduced a renormalization $Z$ of the kinetic term that in
the sphaleron case depends on temperature $T$, the field value
$\phi$ and the gauge parameter $\xi$. The coefficient $C(x)$ from
the last section that enters the Nielsen identity for the effective
action depends functionally on the field value. It can be expanded in
field gradients as~\cite{Metaxas:1995ab}
\be
\label{eq:C_grad_ex}
C(x) = C_0(\phi) + D(\phi) \partial_\mu \phi \partial^\mu \phi 
       - \partial^\mu [\tilde  D(\phi) \partial_\mu\phi] + {\cal O}(\partial^4) \, ,
\ee
where the coefficients depend on the field value evaluated at position
$x$.  Note that the term involving $\tilde D$ corresponds to a total
derivative.  Nevertheless it gives rise to a non-zero contribution to
the right-hand side of the Nielsen identity for the effective action,
and it turns out that it has to be taken into account in a consistent
gradient expansion.  This term has been neglected
in~\cite{Metaxas:1995ab} but was not important there at leading order
due to the different counting $\lambda \sim g^4$. The Nielsen identity
is fulfilled when in leading order in gradients the identity
(\ref{eq:newNielsen}) for the effective potential is established,
while in the next order in gradients one finds\footnote{At finite
  temperature the correction to the kinetic term can differ in general
  for the temporal and spatial components. The Nielsen identity can be
  easily generalized to this case. In the following, $Z$ refers to
  the correction corresponding to the spatial components.}
\be
\label{eq:ZC_relation}
\xi \frac{\partial Z}{\partial \xi} = 
- C_0 \frac{\partial Z}{\partial \phi}
- 2 Z \frac{\partial C_0}{\partial \phi}
+ D \frac{\partial V_{eff}}{\partial \phi}
+ \tilde D \frac{\partial^2 V_{eff}}{\partial \phi^2}.
\ee

Notice that this relation cannot ensure the gauge independence of the
tunnel action or the sphaleron rate, since the gradient expansion does
not apply in these cases, due to $\partial_\mu \phi \partial^\mu \phi
\simeq V_{eff}$. Still, this relation constitutes an essential check
of the consistency of the perturbative scheme that is used to evaluate
the effective action. The topic of the subsequent sections is to study
this identity in the Abelian Higgs model close to the broken phase
explicitly, where perturbative evaluation of all quantities is
plausible.

\section{The effective potential to order $g^3$ and $\lambda$
  \label{sec:effpot}}

In the following we reproduce the effective potential to order $g^3$
and $\lambda$ close to the broken phase. Subsequently, we discuss its
gauge dependence and the corresponding Nielsen identity.  Before we do
so, we briefly motivate why a counting $g^3 \sim \lambda$ is the
appropriate choice in the context of cosmology and finite
temperature. It is well known that in the present model the strength
of the phase transition crucially depends on cubic contributions to
the effective potential of the form $(g \phi)^3 T$. At the same time
the strength of the phase transition is for viable
baryogenesis~\cite{Shaposhnikov:1986jp, Anderson:1991zb} constrained
by $\phi_c/T \sim g^3/\lambda > 1$. So the largest value of $\lambda$
that is interesting in cosmology is of order $g^3$.  Besides, for
larger values of $\lambda$, the convergence of perturbation theory
becomes worse and completely breaks down at $\lambda \sim g^2$. For
smaller values of $\lambda$, the convergence of the perturbative
expansion improves. However, the high temperature expansion that we
want to employ in the following would break down since for $\lambda
\sim g^4$, we find $m_W/T \sim g \phi_c /T \sim 1$.  Therefore, we
count in the following $\phi_c \sim T$ and $\lambda \sim g^3$ as was
done in~\cite{Arnold:1992rz}
\footnote{At zero temperature, the counting $\lambda\sim g^4$ is more
  appropriate~\cite{Metaxas:1995ab, Coleman:1973jx}.}.

At one-loop order the effective potential is given by
\be
V_{eff}(\phi) = V(\phi) + \frac12 T^4 \sum_i \, n_i \,  I_B(m_i^2/T^2) \, ,
\ee
where the sum runs over all species, $n_i$ depends on the statistic of
the field, and the function $I_B$ is given by
\bea
\label{eq:JBdef}
I_B(y) &=& \frac{1}{2 \pi^2} \sum_n \int dx \, x^2 \, 
\log{(4\pi^2 n^2 + x^2 + y)} \nn \\
&\simeq& \textrm{const}\,  + \, \frac1{12} y - \frac{1}{6\pi} y^{3/2} + {\cal O}(y^2) \, . 
\eea
Notice that the zero temperature contributions are of order $y^2$ and
are neglected in the following.

The tree level inverse propagators of the Higgs, Goldstone, gauge
fields and ghosts are respectively
\bea
i P^{-1}_{h/\chi/FP} &=& p^2 - m_{h/\chi/FP}^2 \,, \nn \\
i P^{-1}_A &=& p^2 g_{\mu\nu} - (1 - 1/\xi) p_\mu p_\nu - m_A^2 g_{\mu\nu} \,, \nn
\eea
where we have introduced the field-dependent masses
\bea
m_h^2 &=& \frac{\lambda}{2}(3 \phi^2 - v^2 ) \, ,  \nn \\
m_\chi^2 &=& \frac{\lambda}{2}(\phi^2 - v^2) - 2\xi g^2 \phi^2\, , \nn \\
m_A^2 &=& 2g^2\phi^2\, , \nn \\
m_{FP}^2 &=& 2\xi g^2 \phi^2 \,.
\eea
These expressions can be used to write the effective potential in the
mean field approximation (using the leading terms in (\ref{eq:JBdef}))
\bea
\label{eq:Vmeanfield}
V_{mean-field} &=& \frac{\lambda}{4} \left( \phi^2 - v^2 \right)^2
+  \frac{T^2}{24} \left( \frac{\lambda}{2}(3 \phi^2 - v^2) +  
\left[ \frac{\lambda}{2} (\phi^2 - v^2) + 2\xi g^2 \phi^2 \right] \right. \nn \\
&& \left. \quad + \phantom{\frac12} 2\left[ 3 g^2 \phi^2 + g^2 \xi \phi^2 \right]  
- 4 \xi g^2 \phi^2  \right)  \nn \\
&=&  \frac{\lambda}{4} \left( \phi^2 - v^2 \right)^2
+  \frac{T^2}{24} \left( \lambda(2 \phi^2 - v^2) +  6 g^2 \phi^2 \right),
\eea
where the different contributions are from the Higgs, the Goldstone,
the gauge bosons and the ghosts, respectively.

Notice that at tree level and for $\phi=v$, the poles of the
Goldstone, the $k^\mu$ polarization and the ghosts are all situated at
$p^2 = 2\xi g^2 \phi^2$ and the corresponding contributions to the
effective action cancel 
\footnote{ A term that depends on $\xi$ but not on $\phi$
  can be canceled by an appropriate measure for the ghost
  fields~\cite{Kobes:1990dc}.}.  However, at finite temperature the minimum
of the effective action moves away from $\phi = v$ and higher loop
corrections become important that have to be resummed. In the
following we show that these two effects indeed cancel each other in
leading order.

\begin{figure}[t!]
\begin{center}
\includegraphics[width=0.4\textwidth, clip ]{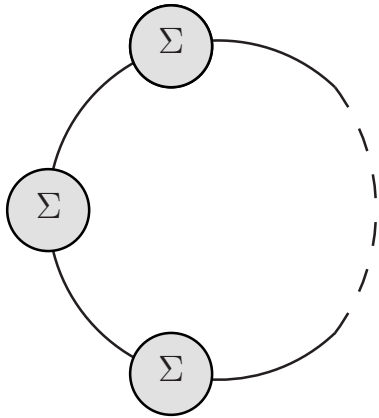}
\end{center}
\caption{
\label{fig:daisies}
\small The daisy diagrams that are resummed.
}
\end{figure}

\begin{figure}[t!]
\begin{center}
\includegraphics[width=0.95\textwidth, clip ]{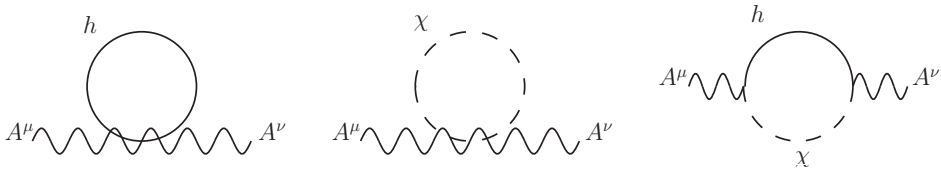}
\end{center}
\caption{
\label{fig:seW}
\small The leading contributions to the self-energy of the gauge bosons.
}
\end{figure}

The necessity of resummation~\cite{Dolan:1973qd, Carrington:1991hz} at
finite temperature arises because of diagrams as depicted in
Fig.~\ref{fig:daisies}.
In general, all finite temperature contributions to the
self-energies are UV finite. Once the sum over the Matsubara
frequencies is performed (or if the real time formalism is used), the
integrand contains the particle distribution functions that are
exponentially suppressed for momenta larger than the
temperature. Hence, the graphs that are apparently UV divergent can be
estimated to be of order of the temperature. In particular, tadpole
diagrams of the self-energy that arise from the gauge interaction are
of order $g^2 T^2$ (e.g.~the contributions to the self-energy of the
gauge bosons shown in Fig.~\ref{fig:seW}).

If the particle in the loop has a mass $m_l$ and the self-energy is of
order $g^2 T^2$, adding self-energies leads to additional factors
\be \frac{g^2 T^2}{(2 \pi n T)^2 + p^2 + m_l^2} \, . \ee 
As long as $n>0$, this yields only a subleading correction of order
$g^2$. Still, for the zero mode $n=0$, this contribution can be
sizable and the according diagrams have to be resummed. This problem
is particularly severe when $m_l^2 \ll g^2 T^2$ and the convergence of
the perturbative expansion crucially depends on this resummation. This
is for example the case for the longitudinal gauge bosons, that in the
symmetric phase are massless on tree level, but receive self-energy
corrections of order $\Sigma \sim g^2 T^2$ at one-loop. But even in
the broken phase the self-energy and the mass are both of order $g^2
\phi^2$ and resummation is essential.

In order to resum these diagrams consistently, one has to absorb at
least a part $\Sigma_0$ of the self-energy into the propagator. In the
broken phase, the details how this is done do not really matter as
long as
\be
m^2_{eff} = m_l^2 + \Sigma_0 \gg \Sigma - \Sigma_0 \, .
\ee
This ensures that IR divergences are tamed and the perturbative series
converges. These self-energy contributions depend on the temperature
and they have to be compensated by introducing a counterterm of equal
size in order to avoid a temperature-dependent regularization
scheme. These counterterms contribute at higher loop level as we
will see below.  $\Sigma_0$ can in principle depend on momentum and
also on the Matsubara number $n$. A particularly simple choice is to
only absorb the self-energy of the zero mode in the
propagator~\cite{Arnold:1992rz}. 

For the choice of resummation scheme, it is crucial that in the present
context we are only interested in static quantities. If non-static
quantities as the plasmon damping rate are considered, the
resummation of the zero modes is not enough to ensure the convergence
of perturbation theory as discussed in~\cite{Kraemmer:1994az}. In this
case more elaborate schemes like hard thermal loop resummation have to
be employed~\cite{Braaten:1989mz}.

In the broken phase (and for $\xi \sim 1$), it suffices to absorb the
leading term of the self-energy of the gauge bosons. After this is
done, the pole structure of all gauge boson propagators is proper in
leading order and corrections can be treated perturbatively. On the
other hand, the symmetric phase is more problematic because the
transverse polarizations remain massless in perturbative calculations.

Within the context of non-Abelian gauge theories, adding additional
lines using the self-interaction vertex of the gauge bosons results in
a factor $g^2 T/m_A$ what is of order $g$ in the broken phase. In the
symmetric phase, on the other hand, the mass of the transverse
polarizations is known~\cite{Braaten:1989mz, Buchmuller:1993bq} to be
non-perturbative and only of order $g^2 T$ what leads to the well
known Linde's problem. For the Abelian case, there is no
self-interaction of gauge bosons and the mass of the transverse
polarization should vanish in the symmetric phase
\cite{Buchmuller:1995xm}. But also here the convergence of the
perturbative expansion is not obvious
\cite{Hebecker:1995kd,Buchmuller:1993bq}.

Compared to the unresummed one-loop result in Eq.~(\ref{eq:JBdef})
resumming only the zero modes gives an additional contribution
\be
T \int \frac{d^3 p}{(2\pi)^3} \log \frac{p^2 + m^2 + \Sigma_0}{p^2 + m^2} \, .
\ee
One way to evaluate this expression is to determine
its derivative with respect to $m^2$ which leads to
\bea
&& \hskip -1 cm T \, \int \frac{d^3 p}{(2\pi)^3} 
\left ( \frac{1}{p^2 + m^2 + \Sigma_0} - \frac{1}{p^2 + m^2} \right) \nn \\
&=& - \frac{T}{4\pi} \left( \sqrt{m^2 + \Sigma_0} - \sqrt{m^2} \right) \, .
\eea
Therefore, the resummation of the zero mode only affects the cubic
term in Eq.~(\ref{eq:JBdef}). The resummed gauge boson masses are given by
\bea
m_T^2 &=& 2g^2\phi^2 \,, \nn\\
m_L^2 &=& 2g^2\phi^2 + a^2g^2\phi^2\,.
\eea
Here, we parametrized the thermal mass of the longitudinal gauge boson
with the parameter $a$ because it depends e.g.~on the fermionic matter
content of the model and is quite different in the Standard Model than
in our example calculation.

Now consider the Goldstone bosons that seem to be more interesting
than the gauge fields due to the $\xi$ dependence in their mass. The
Goldstone bosons are massless on tree level for $\xi=0$, such that
resummation can have a large impact on them.  If the leading high
temperature contributions are taken into account, the self-energy
arises from the diagrams in Fig.~2 and the effective mass is given by
\begin{figure}[t!]
\begin{center}
\includegraphics[width=0.75\textwidth, clip ]{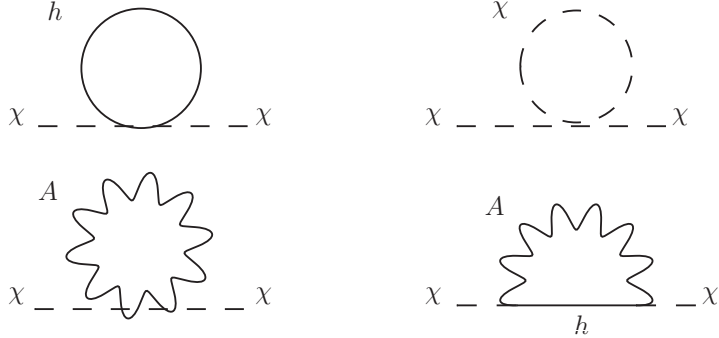}
\end{center}
\caption{
\label{fig:seG}
\small The leading contributions to the self-energy of the Goldstone
bosons.  }
\end{figure}
\be
\textrm{order} \, g^2: \quad \quad
m^2_{\chi, eff} = \frac{\lambda}{2} (\phi^2 - v^2) + 2\xi g^2 \phi^2 
+ \frac{T^2}{12} (\lambda + 3 g^2).
\ee
This fits nicely with the mean field result in
Eq.~(\ref{eq:Vmeanfield}). If our counting was $\lambda \sim g^2$, the
mean field result would give the correct leading order result for the
effective potential and the above expression shows that the Goldstone
boson has a mass $2\xi g^2 \phi^2$ to this order.
This ensures that the cubic terms of the Goldstone boson
and the ghost cancel in the broken phase once the next to leading
order is taken into account.  However, our counting is $\lambda \sim
g^3$, and to leading order the cubic terms coming from the three
physical gauge bosons are important.  Even though this term is
$\xi$-independent, it induces an additional shift in the vev such that
the masses of the Goldstone boson and the ghosts do not coincide any
more in the broken minimum of the potential. This problem is particularly
severe for small $\xi$, since the ghost becomes massless while the
Goldstone boson has a mass of order
\be
m_{\chi, eff}^2 \sim \frac{1}{\phi \, d\phi}\, T \, m_A^3 \sim g^3 T^2,
\ee
so the cubic contribution from the Goldstone boson is (partially)
screened while the contribution from the remaining ghost is not. This
introduces a gauge dependence of order $g^3$ into the effective
potential.

\begin{figure}[t!]
\begin{center}
\includegraphics[width=0.4\textwidth, clip ]{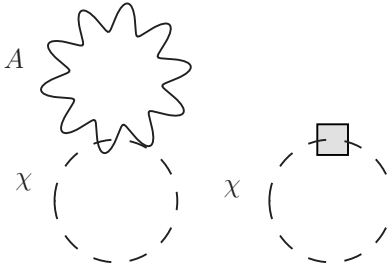}
\end{center}
\caption{
\label{fig:twoloop}
\small Two-loop contribution to the effective potential. The box
denotes the counterterm of the thermal mass.}
\end{figure}

The solution to this dilemma is that there is a subleading
contribution from the gauge bosons to the self-energy of the Goldstone
bosons. This is easily seen by realizing that the integral that arises
in the tadpole contribution
\be
J_B(y) = \frac{1}{2\pi^2} \sum_n \int dx x^2 \frac{1}{4 \pi^2 n^2 + x^2 + y} \, ,
\ee
is nothing else than the derivative of the one-loop vacuum
contribution in Eq.~(\ref{eq:JBdef}) 
\be
J_B(y) = \frac{d}{dy} I_B(y) \, .  
\ee
Hence there is a contribution from the two-loop
diagrams~\cite{Arnold:1992rz} depicted in Fig.~\ref{fig:twoloop} that
is of order $g^2 m_A T^3 \sim g^3 T^3 \phi$.
If $\xi \lesssim g$ this contribution needs to be resummed.  This
then ensures that the Goldstone boson has the same mass as the ghost in the
minimum of the potential to order $g^3$ and their cubic contributions
cancel each other once they are taken into account. The resummed
Goldstone mass is given by
\bea\label{eq:GB}
\textrm{order} \, g^3: \quad \quad
m^2_{\chi, eff} & = & \frac{\lambda}{2} (\phi^2 - v^2) + 2\xi g^2 \phi^2
  + \frac{T^2}{12} (\lambda + 3 g^2) \\
&& - \frac{Tg^2}{4\pi}\left( 2 m_T \nn
+ m_L \right).
\eea
In the following, we denote this expression for the Goldstone mass by
$m^2_\chi$.

\begin{figure}[t!]
\begin{center}
\includegraphics[width=0.4\textwidth, clip ]{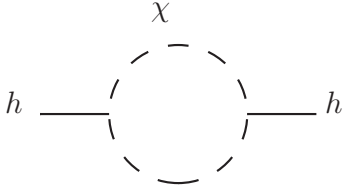}
\end{center}
\caption{
\label{fig:seh}
\small The Goldstone contributions to the self-energy of the Higgs
boson.  }
\end{figure}

Finally, we comment on the contribution from the Higgs. Because the
Higgs has a mass of order $\lambda \phi^2$, its cubic contribution
and thermal mass are not relevant to our analysis. Still, there is a
subtlety coming from the fact that the loop contribution to the
self-energy in Fig.~\ref{fig:seh} is of order $\lambda^2 \phi^2
T/m_\chi$ and diverges in the limit $\xi\to 0$. This divergence is due
to the fact that the self-energy is evaluated at vanishing external
momentum. If it is evaluated with external momenta close to the Higgs
mass, it is of order $\lambda^2 \phi^2 T/k \sim \lambda^{3/2} T \phi$
what is small compared to the tree level mass and no resummation is
necessary. Anyway, the numerical impact of this contribution is very small
as long as $\xi$ is not strictly zero.

In conclusion, we arrive to the following expression for the effective
potential to order $g^3$ and $\lambda$
\bea
\label{eq:Vg3}
V_{(g^3, \lambda)} &=& \frac{\lambda}{4} \left( \phi^2 - v^2 \right)^2
+  \frac{T^2}{24} \left( \lambda(2 \phi^2 - v^2) +  6 g^2 \phi^2 \right) \nn \\
&& -\frac{T}{12\pi} \left( 2 m_T^3 
+ m_L^3 +  m_{\chi}^3 - m_{FP}^3 \right) \,.
\eea
We emphasize that the resummed mass of the Goldstone boson is given by
(\ref{eq:GB}). As argued before, the contribution from the Goldstone boson plus ghost is
small when evaluated close to the minimum of the potential. For small
$\xi \ll \lambda/g^2$ it is of order $\lambda^{3/2}$, while for $\xi
\sim 1$ one finds
\be
\label{eq:del_V_shift}
\delta V \sim  -\frac{T}{12\pi} \left(  (m^{2}_{\chi})^{3/2} - (m_{FP}^2)^{3/2} \right)
\sim -\frac{T m_{FP}}{8\pi} (m^2_{\chi} - m_{FP}^2) \, ,
\ee
which is of order $g \lambda \sim g^4$ and also subleading.  However,
if it is not neglected, it induces a slight shift in the minimum
of order 
\be
\label{eq:shift_phi}
\delta \phi \simeq \frac{ \delta V^{'}}{V^{''}}
\simeq g \sqrt{\xi} \frac{ T}{8\sqrt{2}\pi} \, .
\ee
A shift in the position of the minimum is expected on general grounds,
because the field value is not a physical observable. Note that close
to the critical temperature this shift is sub-leading,
\be
\delta\phi/\phi_c\sim g\sqrt{\xi} \,,
\ee
because in our counting $\phi_c/T_c\sim 1$.  Other contributions to
the potential of order $g \lambda$ could eventually reduce this shift
but actually this shift persists even if higher order contributions
are taken into account. This can be seen by inspecting all two-loop
diagrams. It turns out that even though some contributions are
nominally of order $g^4$, they either do not depend on $\xi$ or are
not linear in the temperature, so formally they cannot cancel
(\ref{eq:del_V_shift}) completely. However, one expects at higher
order a contribution of the form
\be
\frac12 \delta \phi_c^2 V^{''} \simeq \frac{g^2 \xi \lambda}{512 \pi^2} T^2 \phi_c^2 \, ,
\ee
to the potential close to the broken phase that removes the
remaining gauge dependence of the effective action evaluated in the
minimum. That this term is not included in the potential (\ref{eq:Vg3})
introduces a subleading gauge dependence in the critical temperature of order
\be
\frac{\delta T}{T} \simeq \frac{\xi \lambda}{256 \pi^2}\, ,
\ee
what is nominally of order $g^3$ and numerically very small.

Let us now discuss to what extent the potential (\ref{eq:Vg3}) respects
the Nielsen identity (\ref{eq:Nielsen_pot}). First, we show
that the shift (\ref{eq:shift_phi}) in the position of the minimum
is supported by the Nielsen identity. As a function of
the parameter $\xi$ and $\phi$, the potential develops iso-potential
curves, i.e. there is a function $\bar \phi(\phi_0,\xi)$ that
fulfills $\bar \phi(\phi_0, \xi_0)=\phi_0$ and according to
(\ref{eq:Nielsen_pot})
\be
\xi \frac{d}{d\xi} V (\bar \phi, \xi) = \xi  
\frac{d \bar \phi}{d\xi} \frac{\partial}{\partial \phi} V 
+ \xi \frac{\partial}{\partial \xi} V 
= C_0 \, \frac{\partial}{\partial \phi} V 
+ \xi \frac{\partial}{\partial \xi} V 
= 0 \, .
\ee
An explicit expression for the function $C_0$ is given in (\ref{eq:exp_C}).
The leading contribution obtained by evaluating the diagram shown in Fig.~\ref{fig:C}
reads
\be
C_0 \simeq \frac12 g^2 \xi \phi T \int \frac{d^3p}{(2\pi)^3} 
\frac{1}{p^2 - m_\chi^2}\frac{1}{p^2 - m_{FP}^2} \, , 
\ee
that gives
\be\label{eq:C_result}
C_0 \simeq \frac{1}{8\pi} \frac{g^2 \xi \phi T}{m_\chi + m_{FP}}
\simeq g \sqrt{\xi} \frac{ T}{16\sqrt{2}\pi} \, , 
\ee
where the last expression is obtained when evaluating the masses close
to the broken phase. This agrees nicely with our findings in (\ref{eq:shift_phi}) and the
relation $C_0 = \xi  d \bar \phi / d\xi $. Using (\ref{eq:C_result}),
it is also possible to verify explicitly that the Nielsen identity for
the effective potential (\ref{eq:Nielsen_pot}) is indeed satisfied also
away from the broken minimum, up to higher order corrections. Namely,
taking the $\xi$-derivative of the effective potential given in
(\ref{eq:Vg3}) one finds
\bea
\xi \frac{\partial V}{\partial\xi} & = &
- \frac{1}{4\pi} g^2 \xi \phi^2 T (m_\chi-m_{FP}) \,.
\eea
The right-hand side is of higher order close to the minimum where
$m_\chi\approx m_{FP}$, such that the value of the effective potential
in the minimum is $\xi$-independent, up to higher order corrections,
as discussed before. Using that $\frac{\partial V}{\partial\phi}
\simeq 2\phi(m_\chi^2-m_{FP}^2) + \mathcal{O}(g^4,\lambda g)$ in
combination with (\ref{eq:C_result}) then shows that the Nielsen
identity is respected also away from the broken minimum.

\section{Gauge Independence in the Gradient Expansion
  \label{sec:grad_check}}

Let us return to the relation (\ref{eq:ZC_relation}) that ensures the
gauge independence of the effective action in the gradient
expansion. An explicit calculation of the wave function correction
appearing in the derivative expansion of the effective action yields
(see Appendix \ref{sec:kinterm})
\bea\label{eq:Z_general}
Z & = & 1 - \frac{g^2T}{3 \pi } \Bigg( \frac{11}{8 m_T} - \frac{m_T^2}{16m_L^3} 
+ \frac{2}{m_\chi+m_T} - \frac{2}{m_{FP}+m_T} \nn\\
&& {} + \frac{\xi}{m_\chi+m_{FP}} -\frac{7\xi}{16m_{FP}}-\frac{\xi m_{FP}^2}{16m_\chi^3}\left(1+\delta/\xi\right)^2 \Bigg)\,,
\eea
where we have defined
\be
\delta \equiv \frac{\xi\phi}{2 m_{FP}^2}\frac{\partial}{\partial \phi} ( m_\chi^2-m_{FP}^2 ) 
\simeq \frac{\lambda}{4g^2} - \frac{g^2T}{8\pi}\left(\frac{2}{m_T}+\frac{1}{m_L}\right) \;.
\ee

Within our counting, $\delta$ is of order $g$ and hence subleading.
Still, we keep these corrections since unlike two-loop
contributions they are linear in the temperature and (as we shall see)
cancel among themselves.  The $\xi$ dependence of $Z$ arises via the
masses $m_\chi$ and $m_{FP}$, apart from the explicit dependence on
$\xi$. Formally, the $\xi$-dependent terms are of the same order as
the $\xi$-independent ones. In the limit $\xi\to 0$, the wave function
correction agrees with the one obtained in Landau gauge (see appendix
A of \cite{Bodeker:1993kj}).

Note that close to the broken minimum where $m_\chi\approx m_{FP}$,
the leading $\xi$-dependent terms cancel and the $\xi$ dependence
becomes suppressed. This behavior is precisely the one expected from
the Nielsen identity (\ref{eq:ZC_relation}), because close to the
broken minimum all terms on the right-hand side are at most of order
$g^2$. Namely, the term proportional to $\partial C_0/\partial\phi$ is
of higher order because close to the broken minimum $C_0$ becomes
approximately $\phi$-independent. The term involving $C_0\partial
Z/\partial\phi$ is suppressed because both factors are of order $g$,
and the term proportional to the derivative of the effective potential
also has to vanish in the broken minimum by definition. Finally, the
last term is suppressed as well because the Higgs mass is of order
$\lambda\sim g^3$. In order to check explicitly that the Nielsen
identity (\ref{eq:ZC_relation}) is also fulfilled for field
configurations away from the broken minimum, we have computed the
leading contributions to the coefficients $D$ and $\tilde D$
contributing to the derivative expansion (\ref{eq:C_grad_ex}) of the
Nielsen coefficient $C(x)$ for the full effective action. It is
possible to relate these coefficients to Feynman diagrams obtained
from attaching one or two external Higgs field lines to the one-loop
graph  shown in Fig.~\ref{fig:C} that corresponds to the leading
contribution to $C(x)$. The diagrams and the explicit expressions
are shown in Appendix \ref{sec:kinterm}.

\begin{figure}[p!]
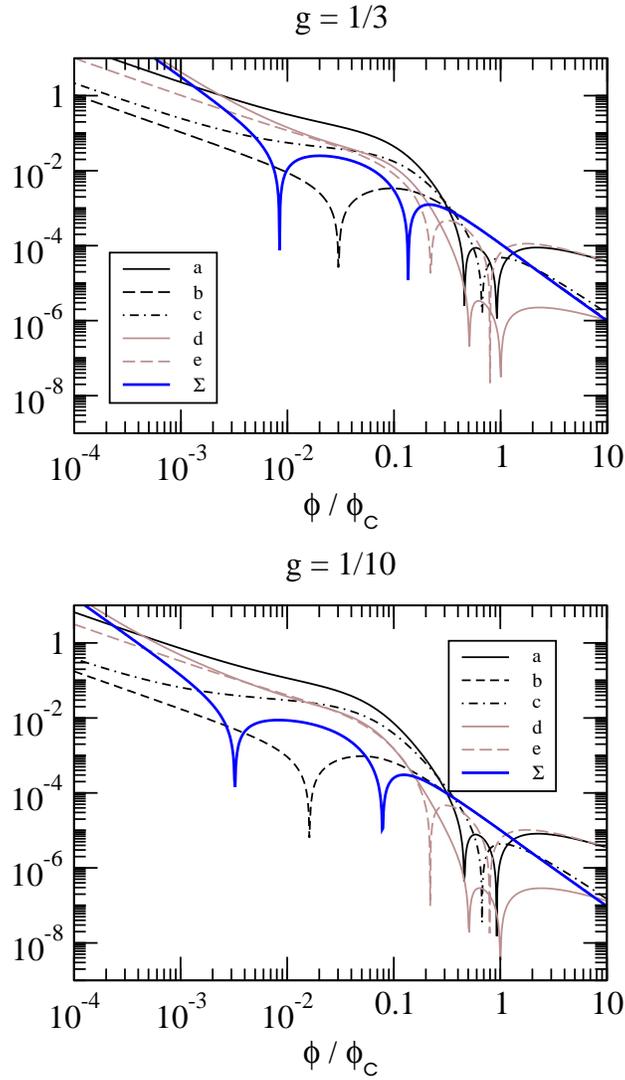

\begin{center}
\includegraphics[width=0.60\textwidth, clip ]{figs/Nielsen_a.eps}
\vskip 0.2 cm 
\includegraphics[width=0.60\textwidth, clip ]{figs/Nielsen_b.eps}
\end{center}
\caption{
\label{fig:Nielsen}
\small The different contributions to the Nielsen identity
(\ref{eq:ZC_relation}) for $Z(\phi)$ and their sum.
The labels (a) to (e) are the five contributions in
(\ref{eq:ZC_relation}) from left to right. The parameters of the
upper plot are $g=1/3$ and $\lambda=0.015$, while the lower uses
$g=1/10$ and $\lambda=4.05 \times 10^{-4}$. This ensures
$\phi_c/T_c \sim 1$ in both cases. The gauge parameter is $\xi=0.5$.
}
\end{figure}

One can check that, with these coefficients (\ref{eq:Dterm}), the
result for $Z$ from Eq.\,(\ref{eq:Z_general}), and $C_0$ from
Eq.\,(\ref{eq:C_result}), the Nielsen identity (\ref{eq:ZC_relation})
for the correction of the kinetic term is indeed satisfied to order
$g^2$.  For this calculation, we found it helpful to express also the
leading contributions to the field derivatives of $C_0$ and $V$ in
terms of the Goldstone and ghost masses, \be \frac{\partial
  C_0}{\partial\phi} \simeq \frac{g^2T\xi
}{8\pi}\frac{m_\chi^2-m_{FP}^2(1+\delta/\xi)}{m_\chi(m_\chi+m_{FP})^2}
\,, \ee and $\partial V/\partial\phi \simeq 2\phi(m_\chi^2-m_{FP}^2)$,
as well as $\partial^2 V/\partial\phi^2 \simeq 2(m_\chi^2-m_{FP}^2) +
4m_{FP}^2\delta/\xi$ up to corrections of order
$\mathcal{O}(g^4,\lambda g)$. Two numerical examples are given in
Fig.~\ref{fig:Nielsen}.

\bigskip

\noindent In conclusion, we have the following findings:

\smallskip

First: The function $C_0$ is close to the broken phase of order $g
\sqrt{\xi} T$ and hence the $\xi$ dependence of the critical field
value $\phi_c$ is subleading. Another important consequence is that
the first-order nature of the phase transition is a gauge-independent
statement~\cite{Nielsen:1975fs, Fukuda:1975di} in the regime where the
perturbative expansion converges
\footnote{In our current scheme this also entails the condition that
  $\xi$ is not much larger than ${\cal O}(1)$.}. 

\smallskip

Second: The Nielsen identity (\ref{eq:ZC_relation}) for $Z$ is in leading order
($g^2$) fulfilled everywhere. The derivative $\partial C_0 / \partial
\phi$ is of order $g^2 T / \phi \sqrt{\xi}$. Hence, according to
(\ref{eq:ZC_relation}) $\xi \partial Z/ \partial \xi$ can be at most
of order $g^2$ if $\partial Z/ \partial \phi$ is at most of order $g$.
From the explicit result (\ref{eq:Z_general}) it turns out that, in
the broken phase, indeed $\xi \partial Z/ \partial \xi$ is of order
$g^2$ while $\partial Z/ \partial \phi$ is of order $g$.  Finally, $D$
is of order $1/g$, so all terms in (\ref{eq:ZC_relation}) are of the
same order and their sum to this order vanishes. 

\smallskip

Third: Even though we cannot check the equation (\ref{eq:ZC_relation})
in full generality to order $g^3$, the one-loop terms are the only
contributions to this order that are linear in the temperature. These
contributions cancel among themselves. The remaining terms in
(\ref{eq:ZC_relation}) to order $g^3$ (denoted by $\Sigma$ in
Fig.~\ref{fig:Nielsen}) result from the contributions proportional to
$T^2$ coming from the product of one-loop terms in $C_0
dZ/d\phi$. These have to cancel against genuine two-loop contributions
that we did not calculate.

\smallskip

Fourth: Close to the symmetric phase the functions $Z$, $D$ and $\tilde
D$ diverge what makes the gauge dependence introduced by subleading
contributions large. However, this does not indicate a breakdown of
perturbation theory. What breaks down is the gradient expansion of the
effective action $\Gamma$ and of the function $C$ that was used to
arrive at the relation (\ref{eq:ZC_relation}). In principle, if the
gradient expansion was avoided, explicitly gauge-independent results
could be obtained. Nevertheless, in practice the gradient expansion is
typically used to determine observables like the sphaleron energy or the
tunneling rate perturbatively. In the next sections, we discuss to what extend this
introduces a gauge dependence to the conventional analysis found in
the literature.

\section{Thermal Tunneling\label{sec:tunnel}}

We start the discussion of thermal tunneling by assuming that the
gradient expansion of the effective action is sound and only later
investigate to what extent this really holds true. The rate of thermal
tunneling is mostly determined by the action of the so-called bounce
solution of the effective action~\cite{Langer:1969bc,
  Coleman:1977py}. The (Euclidean) action with $O(3)$ symmetry is in
leading order given by
\be
\label{eq:Eu_action}
\Gamma = \frac{4 \pi}{T} \, \int d\rho \, \rho^2 \left( 
\partial_\rho \phi \partial_\rho \phi \, Z(\phi)
+ V(\phi) \right).
\ee
To simplify the discussion, we assume that the temperature is close to
the critical temperature where the solutions of the equation of motion
is in the `thin-wall-regime'. In this case, the field only changes in
a region $\rho \sim R \gg L$, where $R$ denotes the size of the
nucleated bubbles while $L$ denotes the wall thickness of the bounce
solution~\cite{Coleman:1977py}. In this approximation, the action can be
rewritten as
\be
\Gamma = \frac{4 \pi}{T} \, R^2 \, \int d\rho \left( 
 ( \partial_\rho \phi )^2 \, Z(\phi)
+ V(\phi) \right) - \frac{4\pi}{3T} R^3 \epsilon \, ,
\ee
where $\epsilon$ denotes the potential difference between the symmetric
and the broken phase. The first integral is invariant under a shift in
$\rho$ leading to a conservation of
\be
 ( \partial_\rho \phi )^2 \, Z(\phi) -  V(\phi) = 
{\rm const} \, .
\ee
For the configuration that dominates the path integral in the
tunneling process the constant on the right side vanishes. This gives
for the action
\be
\Gamma =\frac{ 4 \pi}{T} \sigma R^2 - \frac{4 \pi}{3 T} R^3 \epsilon \, ,
\ee
where we defined the wall tension\footnote{Our definition of the
wall tension differs from the usual definition by a factor $\sqrt{2}$
due to our conventions for the kinetic term.}
\be
\label{eq:wall_tension}
\sigma = \int d\rho \left( 
 ( \partial_\rho \phi )^2 \, Z(\phi)
+ V(\phi) \right) = \int_0^{\phi_c} d\phi \sqrt{ V(\phi) Z(\phi)} \, .
\ee
Finally, the bubble size $R$ can be obtained by extremizing this
expression, what gives the well-known result~\cite{Linde:1980tt}
\be
R = \frac{2 \sigma}{\epsilon}, \quad
\Gamma = \frac{16 \pi}{3 T} \frac{\sigma^3}{\epsilon^2} \, .
\ee

This little exercise shows that a gauge-independent tunneling rate can
in the thin-wall regime only be obtained if the wall tension is
gauge-independent (the potential difference $\epsilon$ is obviously
gauge-independent due to Nielsen's identity). At first sight, it seems
to be impossible that the wall tension is gauge-independent: In the
effective potential the gauge dependence starts at relative order $g$,
while in $Z$ the gauge dependence first occurs at relative order
$g^2$. However this argument is not sound, since the gauge dependence
arising from the potential is further suppressed in the wall
tension. To see this, one can ignore the subleading contributions in
$Z$ to order $g^2$. This gives then for the wall tension
\be
\label{eq:N_wall_tension}
\xi \frac{d}{d\xi} \sigma \simeq \int_0^{\phi_c} d\phi \, \xi \frac{d}{d\xi} \sqrt{ V} 
\simeq  \int_0^{\phi_c} d\phi\, C_0 \frac{d}{d\phi} \sqrt{ V} 
\simeq -  \int_0^{\phi_c} d\phi\, \frac{dC_0}{d\phi} \sqrt{  V} \, . 
\ee
Here we used that the potential vanishes in both phases in the
thin-wall approximation.  $dC_0/d\phi$ is in the broken phase of order
$g^2$ such that the gauge dependence of the effective potential can in
the wall tension cancel against the gauge dependence in $Z$.

Is this result specific to the thin-wall regime and how does this fit
together with the statement (\ref{eq:ZC_relation})?  On first sight,
(\ref{eq:ZC_relation}) only ensure the gauge independence of the
action in the gradient expansion, or more specifically as long as $Z
\partial_\mu \phi \partial^\mu \phi \ll V$. On the other hand, for the
tunneling bounce solution both terms are of the same order and this
expansion does not apply. However, in the present context there is a
way to derive a similar relation as (\ref{eq:ZC_relation}) for the
tunneling bounce without this constraint. Starting from
(\ref{eq:DelW}), a functional derivative with respect to $j(x)$ yields
\be
\label{eq:phi_shift}
\xi \frac{d \phi(x)}{d \xi} = C(x, \phi(x), \xi) \, ,
\ee
in case $\phi(x)$ fulfills the equation of motion, $j(x)=0$. So, there
is a class of solutions $\varphi(x, \xi)$ with
\be
\label{G_indep_eom}
\frac{d}{d\xi} \Gamma(\phi(x), \xi) |_{\phi(x)=\varphi(x, \xi)} = 0 \, .
\ee
Using this relation in (\ref{eq:Eu_action}) yields
\be
0 = 4 \pi T \, \int d\rho \, \rho^2 \left( 
 \partial_\rho \phi \partial_\rho \phi \, 
\left[ 2 \frac{dC}{d\phi} + \frac{dZ}{d\phi} C + \xi \frac{dZ}{d\xi}\right]
+ \xi \frac{dV}{d\xi} + C \frac{dV}{d\phi} \right). 
\ee
So qualitatively the same picture emerges as in the thin-wall
regime. To relative order $g$, the shift (\ref{eq:phi_shift}) in
combination with (\ref{eq:Nielsen_pot}) ensures the gauge independence
but only because $dC/d\phi$ is of order $g^2$ while $C$ is of order
$g$.

Using the expansion (\ref{eq:C_grad_ex}) in (\ref{G_indep_eom}) then
leads to a relation similar to (\ref{eq:ZC_relation}). Unlike the
condition $Z \partial_\mu \phi \partial^\mu \phi \ll V$, the expansion
(\ref{eq:C_grad_ex}) is well justified in the present context at least
in the broken phase. Nominally the expansion parameter is $p^2 / m^2$
such that with $p^2 \phi_c^2 \sim V \sim g^3 \phi_c^4$ and masses of
order $m^2 \sim 2 \xi g^2 \phi^2$ the gradient expansion of
(\ref{eq:C_grad_ex}) is valid for $\phi \gg \sqrt{g/\xi} \phi_c$.
In particular, the gradient expansion gets worse for small $\xi$. 

Now notice that in the last section we have shown the relation $\xi
dV/d\xi = C_0 \, dV/d\phi$ only up to terms of order $g^5$ and that
$dD/d\phi$ is like $dC_0/d\phi$ of order $g^2$. Hence to order $g^5$,
the relation (\ref{G_indep_eom}) is not exactly equivalent to
(\ref{eq:ZC_relation}). In any case, the range of validity is not the
same for both relations, since (\ref{eq:ZC_relation}) is valid for
small gradients, while in the derivation of (\ref{G_indep_eom}) we
used the equation of motion.

From above discussion, it seems that the gauge dependence of the wall
tension is suppressed at least by $g^2$ if $Z$ is ignored and that the
inclusion of the $Z$ factor might even postpone it to order
$g^3$. However, the breakdown of the gradient expansion in the
symmetric phase prevents a gauge-independent determination of the wall
tension in (\ref{eq:wall_tension}) to this order. In practice, the
expression we obtained for the wave function normalization $Z$ can
even become negative for $\phi \sim g \phi_c$ such that no
reasonable results can be obtained if the leading non-trivial order of
$Z$ is taken into account. Also $dC_0/d\phi$ is of order $\sqrt{g}$
close to the symmetric phase what indicates that a large
gauge dependence arises if the gradient expansion is employed.

\begin{figure}[p!]
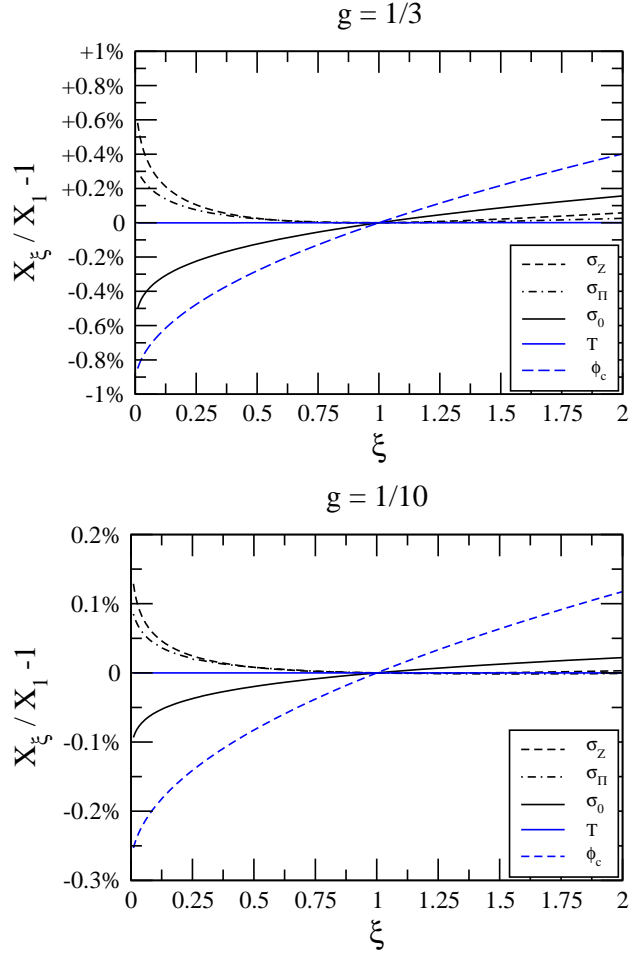

\begin{center}
\includegraphics[width=0.60\textwidth, clip ]{figs/Tunnel_a.eps}
\vskip 0.2 cm 
\includegraphics[width=0.60\textwidth, clip ]{figs/Tunnel_b.eps}
\end{center}
\caption{
\label{fig:Tunnel}
\small Dependence of the wall tension $\sigma$ on the gauge parameter $\xi$.
  The three values $\sigma_0$,
  $\sigma_Z$ and $\sigma_\Pi$ denote the wall tension deduced, from
  $V$ only, from $V$ and the naive $Z$ and from the full momentum
  dependence in $\Pi$, respectively. For comparison, also the critical Temperature $T_c$
and the critical field expectation value $\phi_c$  are shown. The residual gauge dependence
of the wall tension is of relative order $g^{3/2}\ln(\xi)$, while for $T_c$
it is very small, of order $g^3$. In contrast, the $\xi$ dependence of $\phi_c$ is
expected from the Nielsen identity, see Eq.~(\ref{eq:shift_phi}), and its leading contribution
scales as $g\sqrt{\xi}$. The parameters of the upper plot are $g=1/3$ and
$\lambda=0.015$, while the lower uses $g=1/10$ and $\lambda=4.05
\times 10^{-4}$.
}
\end{figure}

The gauge dependence arises mostly from the symmetric phase and
integration of Eq.~(\ref{eq:N_wall_tension}) leads to the estimate
\be
\label{eq:wall_gauge_dep}
\xi \frac{d\sigma}{d\xi} \simeq \frac{1}{48 \pi} m_\chi^2 T_c \, , 
\ee
for $\xi$ not too small. Here $m_\chi$ denotes the Goldstone mass
for $\phi=0$, that coincides with the Higgs mass $m_h$ in the symmetric
phase. Compared with the leading order result
\be
\sigma \simeq \frac{1}6 m_\chi \phi_c^2 \, ,
\ee
the uncertainty from the gauge dependence scales as $\sqrt{\lambda}\sim g^{3/2}$,
\be
\frac{\xi}{\sigma} \frac{d\sigma}{d\xi} \simeq 
 \frac{1}{8 \pi} \frac{m_\chi T_c}{\phi_c^2} \, . 
\ee
This leads to a logarithmic dependence on $\xi$,
\be
\sigma(\xi)/\sigma(1)-1\propto g^{3/2}\ln(\xi) \, ,
\ee
which nicely fits with our numerical findings shown in Fig.~\ref{fig:Tunnel}.
Especially, this means that for very small values of $\xi$ the gauge dependence
becomes enhanced. For $\xi\lesssim g$, the logarithm is ultimately cut off,
and the $\xi$ dependence is of the relative order $g^{3/2}\ln(g)$. Therefore,
it seems that a choice $\xi\gg{O}(g)$ is slightly preferable for the computation of
the wall tension compared to e.g. Landau gauge.

Including the wave function correction $Z$ to the kinetic term in the
determination of the wall tension mainly leads to a gauge-independent
shift. This shift can be estimated as (details are again given in
Appendix \ref{sec:walltension})
\be
\label{eq:sigma_xi_dep}
\Delta \sigma \simeq -\frac{11\sqrt{2}}{192\pi} g m_\chi T_c \phi_c \, . 
\ee
This scales as $\Delta \sigma/\sigma \sim g$ what is parametrically
larger than the gauge-dependent contributions to the wall tension that
scale as $g^{3/2}$.  The gauge dependence of the wall tension is not
improved systematically by the inclusion of $Z$. Especially, it is
still of order $g^{3/2}$. Nevertheless it turns out that a
cancellation of the $\xi$-dependent contributions occurs for $\xi$ of
order one.  These findings are supported by our numerical results that
are presented in Table~\ref{tab:tunnel} and
Fig.~\ref{fig:Tunnel}. Some analytical details of this discussion are
given in Appendix~\ref{sec:walltension}.

At this point, we would like to comment on~\cite{Kripfganz:1995qi}
that compared the wall tension in the Landau and Feynman gauges for an
effective three dimensional theory. There it was also found that including the $Z$
factor does not lead to a systematic reduction of the
gauge dependence. Nevertheless, the analysis showed that the
gauge-independent contributions from $Z$ where as large as the
gauge-dependent ones. This discrepancy compared to our analysis is due
to the fact that the Goldstone mass was not resummed what leads to a
larger gauge dependence in the effective potential as stressed in
section~\ref{sec:effpot}.

The fact that the inclusion of $Z$ does not postpone the gauge
dependence of the wall tension to the relative order $g^2$ or even
$g^3$ is related to the breakdown of the gradient expansion in the
symmetric phase, which leads to a divergence of $Z$ for $\phi\to 0$.
To avoid this problem, we would like to check the gauge independence
of the effective action in the symmetric phase without resorting to
the gradient expansion. In order to achieve that, one can expand the
effective action in $\phi$ around the symmetric phase and obtains
\be
\Gamma = T \int \frac{d^3p}{(2\pi)^3} \, \phi(p) \Pi(p) \phi(p) \, ,
\ee
with
\be
\Pi (p) =  p^2 +  m_h^2  + \Pi_2 (p) \, .
\ee
The gauge independence of the effective action (\ref{eq:Nielsen}) then
implies in leading order
\be
\label{eq:Pi_Nielsen}
\xi \frac{d \Pi_2(p)}{d \xi} = 2 (p^2 +   m_h^2) \frac{ dC(p)}{d\phi} \, ,
\ee
where $C(p)$ is now understood to be expanded in $\phi$ instead of
$p$. In the limit $\phi \to 0$, the only contribution at one-loop
order to $\Pi_2$ is the last diagram depicted in Fig.~\ref{fig:self_Z}
involving the gauge and Goldstone boson and several tadpole diagrams
that however have no momentum dependence. Explicit calculation shows
\be
\xi \frac{d\Pi_2}{d\xi} = \frac{g^2 T \xi}{4 \pi}
\left[ -  m_\chi + \frac{2(p^2 +  m_\chi^2)}{p} {\rm arctan} ( p / m_\chi) \right] \, ,
\ee
and $dC/d\phi$ in accordance with (\ref{eq:Pi_Nielsen}). Notice also
that $\Pi_2$ is finite in the limit $p \to 0$ if the limit $\phi \to
0$ is taken first
\be
\xi \frac{d\Pi_2}{d\xi} \simeq \frac{g^2 T \xi}{4 \pi}
\left[  m_\chi + \frac83 \frac{p^2}{ m_\chi} \right] \, .
\ee
The mass term agrees hereby with the one derived from the potential
(\ref{eq:Vg3}), while the kinetic term does not
\be
Z_{\rm symm} = 1 - \left.\frac{\partial\Pi_2}{\partial p^2}\right|_{p=0} 
= 1 - \frac{g^2 T}{3 \pi  m_\chi} (2 + \xi) \, .
\ee
Since the potential agrees independent from what limit ($\phi \to 0$
or $p \to 0$) is taken first, it is tempting to reduce the gauge
dependence of the wall tension by interpolating between the two
different kinetic terms $Z$ and $Z_{\rm symm}$. In order to implement
this idea, we replace the wave function correction in Eq.~(\ref{eq:wall_tension})
by
\be
Z(\phi,p) =  1 - \frac{\partial\Pi_2(\phi,p)}{\partial p^2}\;.
\ee
Here $\Pi_2(\phi,p)$ is the full one-loop self-energy for a general expectation
value $\phi$ of the Higgs field, which approaches
the self-energy $\Pi_2(p)$ discussed above for $\phi\to 0$. On the other hand, 
$Z(\phi,p\to 0)$ agrees with the correction to the kinetic term from
Eq.~(\ref{eq:Z_general}). For any non-zero value of the momentum $p$,
$Z(\phi,p)$ has a regular behavior for $\phi\to 0$, and indeed interpolates
between $Z$ and $Z_{\rm symm}$ as a function of $\phi$ provided
that the momentum is chosen small enough, $p \lesssim m_h$. In Table~\ref{tab:tunnel}
we show the resulting expression for the wall tension, where we used $p=m_h/(2\pi)$
as the momentum cut-off. Unfortunately, we found that
qualitatively the gauge dependence did not improve significantly by doing so, although
it is slightly reduced on a quantitative level. A complete cancellation of the gauge
dependence at the relative $g^{3/2}$ level seems to require the use of the full
momentum dependence of the effective action without resorting to the gradient expansion.

\begin{table}
\begin{center}
\begin{tabular}{|c|c|c|c|c|c|c|}
\hline
 & \multicolumn{3}{|c|}{$g = 1/3$} & \multicolumn{3}{|c|}{$g = 2/3$} \\
\hline
 $\lambda$ & 0.0075 &  0.015 & 0.025 & 0.06 & 0.1 & 0.15 \\
\hline
\hline
 $\sigma_0  \, (\xi=1)$ & 6.29 &  1.78 & 0.74 & 149 & 31.2 & 12.9 \\
\hline
 $\delta \sigma_0$ & 0.11\% &  0.33\% & 0.79\% & 0.33\% & 0.70\% & 1.4\% \\
\hline
\hline
 $\sigma_Z \, (\xi=1)$  & 6.16 &  1.70 & 0.68 & 143 & 28.8 & 11.3 \\
\hline
 $\delta\sigma_Z$  & -0.10\% &  -0.25\% & -0.56\% & -0.21\% & -0.46\% & -0.97\% \\
\hline
\hline
 $\sigma_\Pi \, (\xi=1)$  & 6.16 &  1.70 & 0.69 & 143 & 28.8 & 11.3 \\
\hline
 $\delta\sigma_\Pi$  & -0.07\% &  -0.16\% & -0.32\% & -0.13\% & -0.25\% & -0.52\% \\
\hline
\end{tabular}
\end{center}
\caption{\label{tab:tunnel} Numerical results for the wall tension in
  units of $10^{-3}v^3$, and the shift $\delta \sigma$ for a change of
  the gauge parameter $\xi$ from $1$ to $0.1$.  The rows $\sigma_0$,
  $\sigma_Z$ and $\sigma_\Pi$ denote the wall tension deduced, from
  $V$ only, from $V$ and the naive $Z$ and from the full momentum
  dependence in $\Pi$, respectively.}
\end{table}
%

\section{Sphaleron numerics\label{sec:sphaleron}}

In this section we briefly discuss the sphaleron
energy following~\cite{Klinkhamer:1984di}. The sphaleron is a static
Higgs-gauge configuration that is a saddle point of the action (which
reflects the energy of the configuration). It has Chern-Simons number
$\frac12$ and is situated half-way between two gauge vacua. Our toy
model does not contain a $SU(2)$ gauge sector and hence no sphaleron
transitions, but the Higgs potential in the Standard Model has
essentially the same features of the Abelian Higgs and it is
reasonable to feed the potential (\ref{eq:Vg3}) into the equations of
motion of the sphaleron to estimate the gauge dependence of the
sphaleron in the Standard Model. In the conventional analysis, the
main difference between the Abelian and the non-Abelian model in terms
of the strength of the phase transition is that there are three times
as many gauge bosons (and ghosts and Goldstones) contributing to the
cubic term hence strengthening the phase transition. We will mimic
that by also presenting results for larger than observed gauge
couplings.

In the non-Abelian case, perturbation theory is plagued by Linde's
problem in the symmetric phase such that the expansion in the coupling
constant becomes questionable for small Higgs vevs. Besides, compared
to the tunneling rate discussed in the last section, the convergence
of the gradient expansion is even more problematic. The gradient
expansion is formally an expansion in the parameter $p^2/m^2$ where
the relevant mass is the one of the gauge bosons. While for the
tunneling bounce this is ${\cal O}(g)$, in the case of the sphaleron
the expansion parameter is ${\cal O}(1)$ and hence not suppressed by
any coupling constant even in the broken phase. However, numerically
the coefficients $D$ and $Z-1$ are slightly smaller than $C_0$ and
$V_{eff}$ (in units of $m_A^2 \phi_c^2$) such that higher orders can
be neglected if this trend continues. This issue is to certain extend
unrelated to the gauge dependence.

At the same time, the sphaleron energy is proportional to $\phi_c$
such that the gauge dependence stemming from $\phi_c$ cannot possibly
be canceled by the gauge dependence of the effective action solely in
the surrounding of the broken phase. In order to quantify our lack of
knowledge on $Z$ and $V_{eff}$ in the symmetric phase, our strategy is
to set $Z$ to $1$ in the numerical analysis and use the
gauge dependence of $V_{eff}$ to estimate the impact of those
contributions to the sphaleron energy.

The differential equations to solve when a spherical Ansatz is used
read 
\bea
\zeta^2 \frac{d^2 f}{d\zeta^2} &=& 
2 f (1-f) (1-2f) - \frac{1}{4} \zeta^2 h^2 (1-f),  \\
\frac{d}{d\zeta} \zeta^2 \frac{dh}{d\zeta} &=& 2 h (1 - f)^2 
+ \frac{1}{g^2} \frac{d V_h}{dh} \, ,
\eea
with the asymptotic behavior 
\be
 f \to \alpha \zeta^2 \, , \quad h \to \beta \zeta \quad 
\textrm{for} \quad \zeta \to 0 \, , \nn 
\ee
and
\be
\label{eq:fh_asympt}
 f \to 1 - \gamma \exp( - \zeta/2) \, , \quad
h \to 1 - \frac{\delta}{\zeta} \exp(- \kappa \zeta) \quad 
\textrm{for} \quad \zeta \to \infty \, ,
\ee
where the parameter $\kappa$ is given by $\kappa^2 = V_h^{''}$ and
the rescaled potential is defined as $V_h(h) = V_{eff}(h \cdot
\phi_c)$ while the rescaled coordinate is $\zeta = g \phi_c |x|$.
\begin{table}
\begin{center}
\begin{tabular}{|c|c|c|c|c|c|c|}
\hline
 & \multicolumn{3}{|c|}{$g = 1/3$} & \multicolumn{3}{|c|}{$g = 2/3$} \\
\hline
 $\lambda$ & 0.0075 &  0.015 & 0.025 & 0.06 & 0.1 & 0.15 \\
\hline
\hline
 $\phi_c/v$ & 0.965 &  0.549 & 0.325 & 2.07 & 1.03 & 0.535 \\
\hline
 $T_c/v$  & 0.454 &  0.575 & 0.721 & 1.00 & 0.94 & 1.05 \\
\hline
\hline
 $\delta \phi_c$ & $0.3\%$ & $0.7\%$  & $1.5\%$ & $0.6\%$ & $1.3\%$ & $3.0\%$ \\
\hline
 $\delta E_{sph}$ & $0.01\%$ &  $0.01\%$  & $0.02\%$ & $0.05\%$ & $0.07\%$ & $0.10\%$ \\
\hline
\end{tabular}
\end{center}
\caption{\label{tab:sphaleron} Numerical results for the sphaleron
  energy. The row $\delta \phi_c$ contains the shift for a change of
  the gauge parameter $\xi$ from $1$ to $0.1$. The row contains the
  corresponding change in sphaleron energy due to a change in the
  shape of the Higgs potential.}
\end{table}

We solve the equations numerically with a shooting algorithm similar
to what is used to find the bounce solution of the tunneling
action. The parameters $\alpha$ and $\beta$ are chosen and the equations
are solved from some position $\zeta_\epsilon$ close to the origin to
a value $\zeta_\omega \sim {\cal O}(10)$. The parameters $\alpha$ and $\beta$
are then varied and we search for simultaneous zeros in the functions
\bea
D_f &=& 1 - f + 2 f^\prime\, , \nn \\
D_h &=& \zeta \kappa ( 1 - h) + (\zeta h)^\prime \, . 
\eea
These two conditions ensure that the numerical solutions smoothly
match the asymptotic behavior given in (\ref{eq:fh_asympt}). The
sphaleron energy is then given by
\bea
E &=& \frac{4 \sqrt{2} \pi \phi_c}{g} \int_0^\infty  d\zeta \, \left[ 
4 \left( \frac{df}{d\zeta} \right)^2
+ \frac{8}{\zeta^2} ( f (1-f) )^2 \right . \nn \\
&& \left. + \frac12 \zeta^2 \left( \frac{dh}{d\zeta} \right)^2
+ (h (1-f))^2 + \zeta^2 g^{-2} V_h \right] \, .
\eea

Some numerical results are given in Table \ref{tab:sphaleron}. Main
impact of the sphaleron energy has the gauge-dependent shift in
$\phi_c$. The gauge dependence in the critical temperature $T_c$ and
the shape of the effective potential are subleading. Moreover, close
to borderline case of sphaleron washout $\phi_c / T_c \sim 1$, the
uncertainty in the sphaleron energy never exceeds a few percent for
Standard Model values of the gauge coupling. In summary, the
uncertainty stemming from the residual gauge dependence is subleading
compared to corrections coming from two-loop contributions to the
effective potential \cite{Fodor:1994bs,Buchmuller:1995sf}.  This
gauge dependence is inherited from the critical vev
\be
\label{eq:N_sphal}
\frac{\xi}{E} \frac{dE}{d\xi} \simeq 
\frac{g \sqrt{\xi}}{16 \pi \sqrt{2}} \frac{T}{\phi_c} \, ,
\ee
and scales as $g$. To remove this gauge dependence one would probably
need to include the next-to-leading order of the kinetic term of the
gauge-bosons. There, a sizable gauge dependence is expected in order
to ensure a gauge-independent position of the pole in the gauge-boson
propagator. Besides, the breakdown of the gradient expansion should
lead to even more severe effects than in the case of thermal tunneling
as discussed above. This will further complicate the determination of
the sphaleron energy with an accuracy beyond the bound
(\ref{eq:N_sphal}).

\section{Summary\label{sec:conclusion}}

Let us summarize our findings concerning the gauge dependence of the
effective action in the Abelian Higgs model in $R_\xi$-gauges. We
explicitly demonstrated various Nielsen identities in the regime where
the use of perturbation theory and the gradient expansion of the
effective action is feasible.

In particular, we have shown that the position of the minimum of the
effective potential transforms in leading order according to
(\ref{eq:Nielsen_pot}) under a change in the gauge fixing parameter
$\xi$. We would like to
emphasize that this result could only be obtained by calculating the
effective potential consistently to order $g^3$ using the counting
$g^3 \sim \lambda$. In particular, this required the resummation of
contributions to the Goldstone boson mass\footnote{At this point of
  the analysis, we depart from \cite{Patel:2011th, Wainwright:2011qy,
    Wainwright:2012zn} that argue for a large gauge dependence of the
  effective action.}  of order $m_A T$ (where $m_A$ denotes
collectively the different masses of the gauge bosons).

Furthermore, we have demonstrated that the off-shell effective action
in gradient expansion transforms according to the Nielsen identity
(\ref{eq:ZC_relation}). However, this relation cannot guarantee the
gauge independence of vacuum transitions for several reasons. First,
the gradient expansion is not well justified in these cases since the
contribution of the kinetic term to the action is of equal size as (or
even larger than) the contribution from the scalar potential. Second,
even for small gradients the gradient expansion and the relation
(\ref{eq:ZC_relation}) break down at some point in the symmetric phase
and vacuum transitions are also sensitive to this regime. Compared to
the analysis at zero temperature \cite{Metaxas:1995ab}, notice also
the additional contribution involving $\tilde D$ that was missing but
also not important in the analysis presented there.

Finally, we discussed the gauge dependence of the tunneling action.
Using the established procedure to calculate the tunneling action
perturbatively (meaning a canonical kinetic term and the appropriate
effective potential), we found for the gauge dependence of the wall
tension the estimate
\be
\label{eq:conc1}
\frac{\xi}{\sigma} \frac{d\sigma}{d\xi} \simeq 
 \frac{1}{8 \pi} \frac{m_\chi T_c}{\phi_c^2} \, ,
\ee
what scales as $\sqrt{\lambda}\sim g^{3/2}$ (the Goldstone mass $m_\chi$ is evaluated in
the symmetric phase). Including corrections to the kinetic term $Z$
leads to corrections of the same order but the gauge dependence is not
persistently reduced. This is due to the fact that the latter
corrections are sensitive to the effective action very close to the
symmetric phase where the gradient expansion breaks down.

Since we did not arrive at an explicitly gauge-independent result for
$\sigma$, this leaves the question what is the best gauge to
chose. Our results develop the strongest $\xi$ dependence for $\xi\sim
0$. For $\xi \gtrsim g$ the wall tension computed using a canonical
kinetic term depends logarithmically on $\xi$, while when including
the correction to the kinetic term the $\xi$ dependence partially
cancels and $\sigma$ is rather insensitive to $\xi$ for $\xi \sim
1$. This could indicate that a $\xi$ value of this order is the
appropriate choice and not Landau gauge (that is mostly used in the
literature and is reproduced in our case by $\xi \to 0$).
Nevertheless, quantitatively the dependence on $\xi$ is rather small
and including the two-loop contributions to the effective potential
will probably have a larger impact in most models (e.g.~contributions
from the gluons~\cite{Bagnasco:1992tx}). We also identified
gauge-independent corrections to the wall tension that arise from the
kinetic term $Z$ and scale as $g$. Also these contributions are more
important than the gauge-dependent ones.

Several non-perturbative studies of tunneling in the Standard Model were presented in refs.~\cite{Kajantie:1995kf, tension_lattice} and a detailed comparison with the perturbative results in Landau gauge was given in~\cite{Moore:2000jw}. Also there it was concluded that the corrections to the kinetic term and two-loop contributions to the effective potential are important to achieve a good agreement with the non-perturbative results for the wall tension. Numerically, these corrections are far more important than the gauge fixing dependence we discussed here.

In case of the sphaleron, the convergence of the gradient expansion is
also problematic. As a naive estimate of the gauge dependence of the
sphaleron energy, we obtained
\be
\label{eq:conc2}
\frac{\xi}{E} \frac{dE}{d\xi} \simeq 
\frac{g \sqrt{\xi}}{16 \pi \sqrt{2}} \frac{T_c}{\phi_c} \, ,
\ee
what is typically of a few percent. Over all, the effective potential
enters in the sphaleron energy mostly via the position of its
minimum. Hence, improving the gauge dependence of the sphaleron energy
will probably require not only to go beyond the gradient
expansion but also to calculate the gauge dependence of the
wave function corrections of the gauge fields, what we did not attempt
here.

The estimates (\ref{eq:conc2}) and (\ref{eq:sigma_xi_dep}) are solely based on the Nielsen identity for the Higgs vev, $C_0 \equiv \xi d \phi /d \xi$. Therefore, the results can be readily carried over to the Standard Model and some of its extensions as the two-Higgs doublet model or singlet extensions. Compared to the Abelian Higgs model, the electroweak sector of these models gives in leading order a gauge fixing dependence of the Higgs vev that is larger by a factor three. 

In conclusion, determining vacuum transitions in an explicitly
gauge-independent fashion is mostly hindered by the breakdown of the
gradient expansion of the effective action (in particular in the
vicinity of the symmetric phase). Still, in the cosmologically most
interesting regime with $g \ll 1$ and $\phi_c / T_c \gtrsim 1$, the
gauge dependence of the tunneling action and the sphaleron energy is
rather small. The situation further improves when the essential cubic
contributions to the effective potential do not solely arise from the
gauge bosons (as in the present Abelian toy model) but e.g.~from
additional degrees of freedom (as in the light stop
scenario~\cite{Carena:1996wj, Delepine:1996vn, Carena:1997ki}) or
from the tree level dynamics of an extended scalar
sector~\cite{Espinosa:2011ax}.

\section*{Acknowledgments}

We thank W.~Buchm\"uller and M.~Laine for helpful discussions.  This
work has been supported by the German Science Foundation (DFG) within
the Collaborative Research Center 676 ``Particles, Strings and the
Early Universe''.

\begin{appendix}
\section{Feynman Rules}

\begin{figure}[t!]
\begin{center}
\includegraphics[width=0.95\textwidth, clip ]{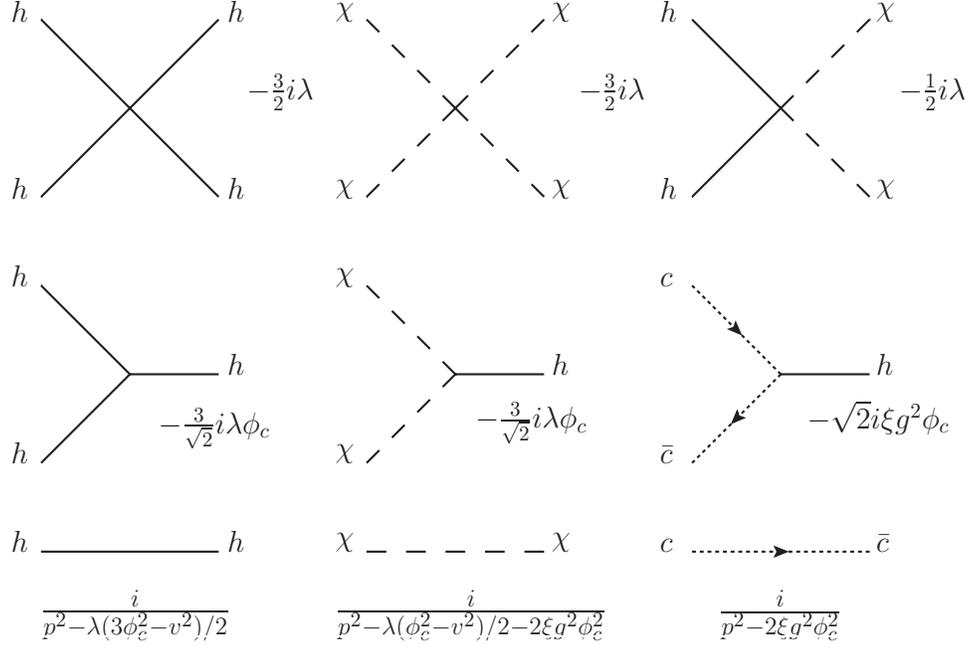}
\end{center}
\caption{
\label{fig:feynman_s}
\small Feynman rules for scalars in the Abelian Higgs model. }
\end{figure}

\begin{figure}[t!]
\begin{center}
\includegraphics[width=0.95\textwidth, clip ]{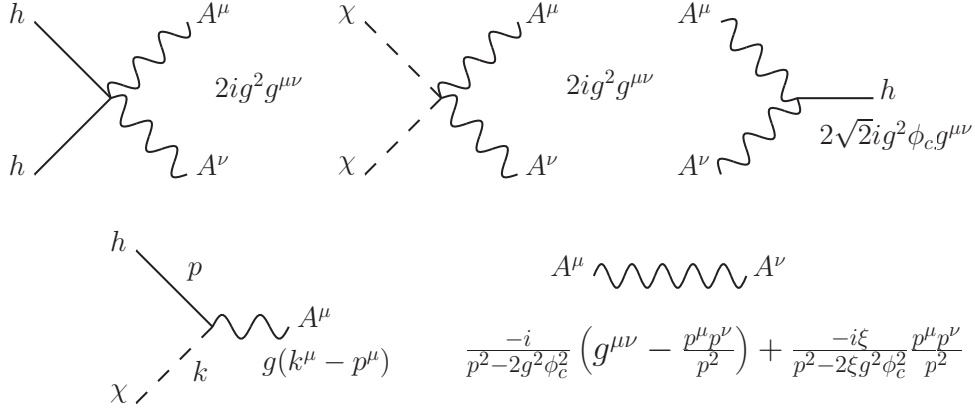}
\end{center}
\caption{
\label{fig:feynman_v}
\small Feynman rules for vectors in the Abelian Higgs model.  }
\end{figure}

The Feynman rules for the scalar particles are shown in
Fig.~\ref{fig:feynman_s} while the Feynman rules involving the vector
particles are given in Fig.~\ref{fig:feynman_v}.
%

\section{Wave function corrections}\label{sec:kinterm}

\begin{figure}[t!]
\begin{center}
\includegraphics[width=0.95\textwidth, clip ]{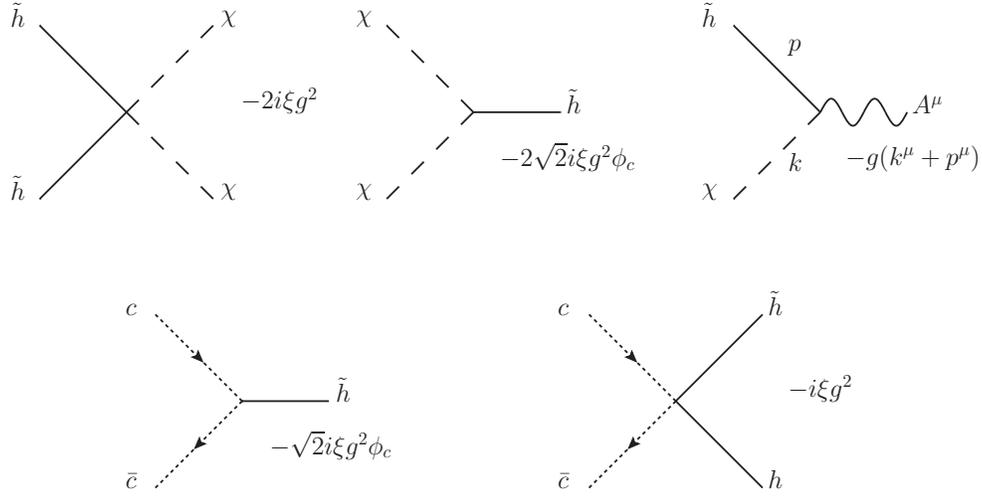}
\end{center}
\caption{
\label{fig:feynman_Z}
\small Additional Feynman rules for the kinetic term
in the effective action. }
\end{figure}

In this section, we present some results on the coefficient of the
kinetic term $Z$ in the $R_\xi$-gauge and also for the functions $D$
and $\tilde D$. As mentioned in section~\ref{sec:model}, derivatives
with respect to the background field $\phi$ are in the $R_\xi$-gauge
not related to diagrams with external Higgs fields.  In order to
remedy this issue, we replace in the gauge fixing terms the background
field $ \tilde\phi$ by $\tilde\phi + \tilde h/\sqrt{2}$ where $\tilde h$ will
be treated as an external field. The function $Z$ is then related to
the two-point functions involving $h$ and $\tilde h$ via
\be 
\label{eq:Zdef}
Z^{\mu\nu} = \frac{\partial^2}{\partial p_\mu \partial p_\nu} 
\left( \Pi_{hh} +  \Pi_{h\tilde h} 
+ \Pi_{\tilde h h} + \Pi_{\tilde h\tilde h} \right). 
\ee
The additional Feynman rules involving the external field $\tilde h$
are shown in Fig.~\ref{fig:feynman_Z}. For the sphaleron and thermal
tunneling, we only need the spatial components $Z^{ii}$ at finite
temperature.

\begin{figure}[t!]
\begin{center}
\includegraphics[width=0.8\textwidth, clip ]{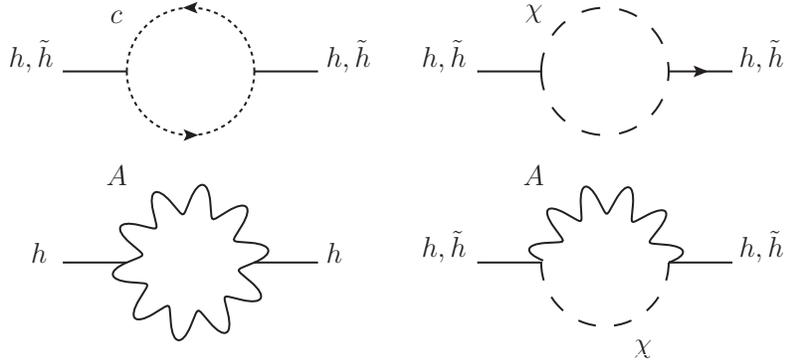}
\end{center}
\caption{
\label{fig:self_Z}
\small The diagrams contributing to the kinetic term of the effective action. }
\end{figure}

On the one loop level the possible diagrams are of the form depicted
in Fig.~\ref{fig:self_Z} where the external lines can be either $h$
or $\tilde h$.
On one loop level it is more practical to first perform the sum in
(\ref{eq:Zdef}) before the integrals are evaluated. The momentum
dependence stems from the integrands
\bea
\label{eq:integrands}
\frac{1}{(l+p/2)^2 + m^2}\frac{1}{(l-p/2)^2 + m^2}
&& \nn \\
&& \hskip -3 cm 
\simeq \frac{1}{(l^2 + m^2)^2} 
- \frac12 \frac{p^2}{(l^2 + m^2)^3}
+ \frac{(p \cdot l)^2}{(l^2 + m^2)^4} \nn \\
&& \hskip -3 cm 
\simeq \frac{1}{(l^2 + m^2)^2} 
- \frac16 \frac{p^2}{(l^2 + m^2)^3}
- \frac13 \frac{p^2 m^2}{(l^2 + m^2)^4} \, ,
\eea
that can be evaluated using
\bea
L_B(y) &=& \frac{1}{2\pi^2} \sum_n \int dx \, x^2 \, 
\frac{1}{(4 \pi^2 n^2 + x^2 + y)^3} \nn \\
&=&  \frac{1}{2} \frac{\partial^2}{\partial^2 y} J_B(y) 
\simeq \frac{1}{32 \pi} y^{-3/2}\, ,
\eea
and a similar expression for the the integrals with higher powers.
Notice that only the zero mode $n=0$ contributes to the leading term
and we neglect higher Matsubara modes in the following whenever
this holds true.

We first restrict ourselves to the broken phase where the masses are
given as
\be
m_A^2 =2 g^2 \phi^2 \equiv m^2 \equiv y T^2 \, , 
\quad m^2_\chi = m_{FP}^2 =  \xi m^2 \, .
\ee
Then one finds for the ghost and Goldstone loops the
contributions
\be
\label{eq:Z_chi2_ghost2}
\frac13 g^2 \xi^2 y (L_B( \xi y) + 2 \xi y M_B( \xi y)) \, .
\ee
The diagram involving two gauge fields requires the evaluation of
products of the polarization projections $P^{\mu\nu} (l) = l^\mu
l^\nu/l^2$
\bea
P^{\mu\nu}(l+p/2) P_{\mu\nu}(l-p/2) 
&\simeq& 1 - \frac{p^2}{l^2} + \frac{(p\cdot l)^2}{l^4} + {\cal O}(p^3) \nn \\
&\simeq& 1 - \frac23 \frac{p^2}{l^2}  + {\cal O}(p^3) \, .
\eea
and again the expansion of the integrand (\ref{eq:integrands}). The
novel integrals are given in appendix \ref{sec:ints}.  The contribution from the
term involving the $p^\mu$ polarization twice gives
\be
\label{eq:Z_mumu}
 \frac13 g^2 \xi^2 y \left( - L_B( \xi y) 
- 2 \xi y M_B( \xi y) \right. 
 \left. - 4 \bar L_B ( \xi y,  \xi y, 0) \right).
\ee
The contribution involving one $p^\mu$ polarization yields
\be
\label{eq:Z_mu}
\frac{8}3 g^2 \xi y \bar L_B ( \xi y, y, 0) \, ,
\ee
and the contribution with no $p^\mu$ polarization is
\be
\label{eq:Z_nomu}
2 g^2 y \left( - \frac12 L_B(y) - y M_B (y)
- \frac23 \bar L_B (y,y, 0) \right).
\ee
Finally, the contributions from the mixed gauge-Goldstone loop give
\be
\label{eq:Z_chimu}
4 g^2 \xi \left( \frac13 \bar K_B( \xi y,  \xi y)\right) \, ,
\ee
and
\be
\label{eq:Z_chimu2}
4 g^2 \left( \frac23 \bar K_B( \xi y, y)\right) \, .
\ee

In the broken phase, where $m^2_{FP}=m^2_\chi$, the contributions
proportional to $L_B$ in (\ref{eq:Z_chi2_ghost2}) and
(\ref{eq:Z_mumu}) cancel each other. The third term involving $\bar
L_B$ in (\ref{eq:Z_mumu}) cancels in leading order against
(\ref{eq:Z_chimu}).  Besides, the terms in (\ref{eq:Z_mu}) and
(\ref{eq:Z_chimu2}) combine to
\be
\frac83 g^2 \left( \xi y \, \bar L_B(\xi y, y, 0) + 
\bar K_B(\xi y, y) \right) \simeq \frac{2 g^2}{3\pi \sqrt{y}} \, ,
\ee
what is gauge-independent. Together with the contribution in
(\ref{eq:Z_nomu}) this yields the final result
\be
Z \simeq 1 - \frac{7 g^2}{16 \pi \sqrt{y}} \, .
\ee
Taking in addition the resummed gauge boson propagator
\be
\frac{-i}{p^2-m_L^2}P_{\mu\nu}^L+\frac{-i}{p^2-m_T^2}P_{\mu\nu}^T+\frac{-i}{p^2-m_{FP}^2}P_{\mu\nu}^0 \, ,
\ee
into account, where $P_{\mu\nu}^0=p^\mu p^\nu/p^2$, 
$P_{\mu\nu}^T=g_{\mu\nu}-u_\mu u_\nu-(\tilde p_\mu\tilde p_\nu)/\tilde p^2$,
$P_{\mu\nu}^L=g_{\mu\nu}-(p_\mu p_\nu)/ p^2 - P_{\mu\nu}^T$, $\tilde p_\mu = p_\mu - u_\mu (up)$,
$u_\mu=(1,0,0,0)$, $m_L^2=m_T^2+ag^2T^2$, and $m_T^2=2g^2\phi^2$, we find
\bea
Z & = & 1 - \frac{g^2}{48 \pi \sqrt{y}} \left(22-\frac{m_T^3}{m_L^3}\right) \,.
\eea

The above calculation can be extended to a configuration in field
space away from the broken minimum. In that case one cannot set
the Goldstone mass $m_\chi$ and the mass $m_{FP}$ of the ghost and
of the time-like gauge boson polarization to be equal. In that case
we obtain the result given in Eq.\,(\ref{eq:Z_general})

\begin{figure}[t!]
\begin{center}
\includegraphics[width=0.85\textwidth, clip ]{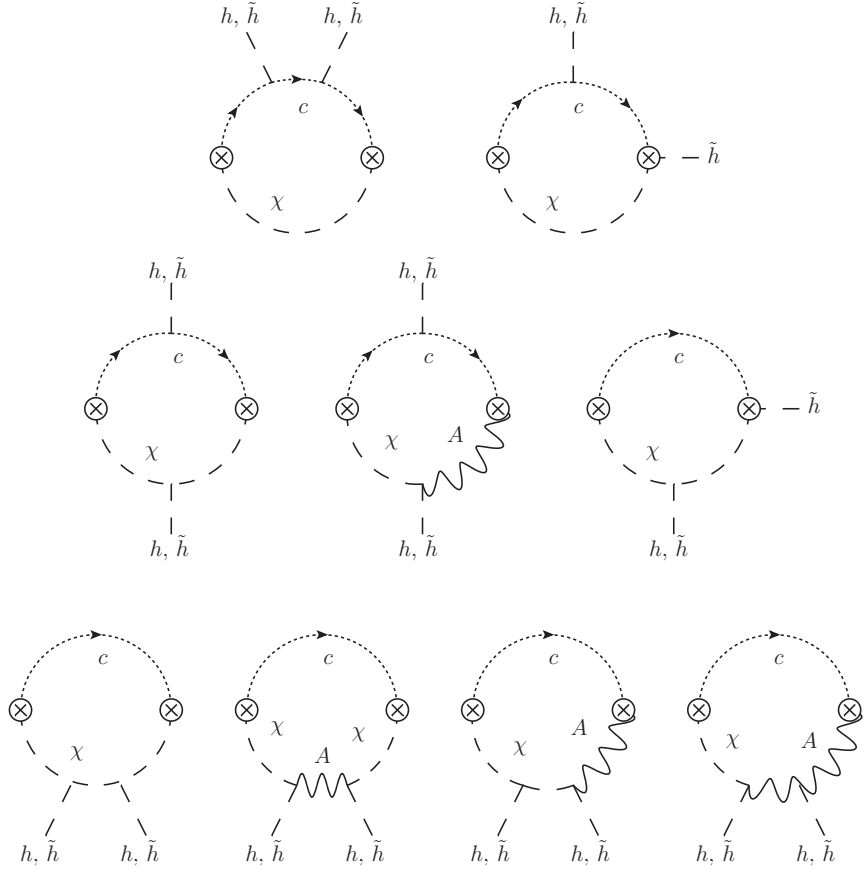}
\end{center}
\caption{
\label{fig:D}
\small The diagrams contributing to the function $D$. }
\end{figure}
\begin{figure}[t!]
\begin{center}
\includegraphics[width=0.6\textwidth, clip ]{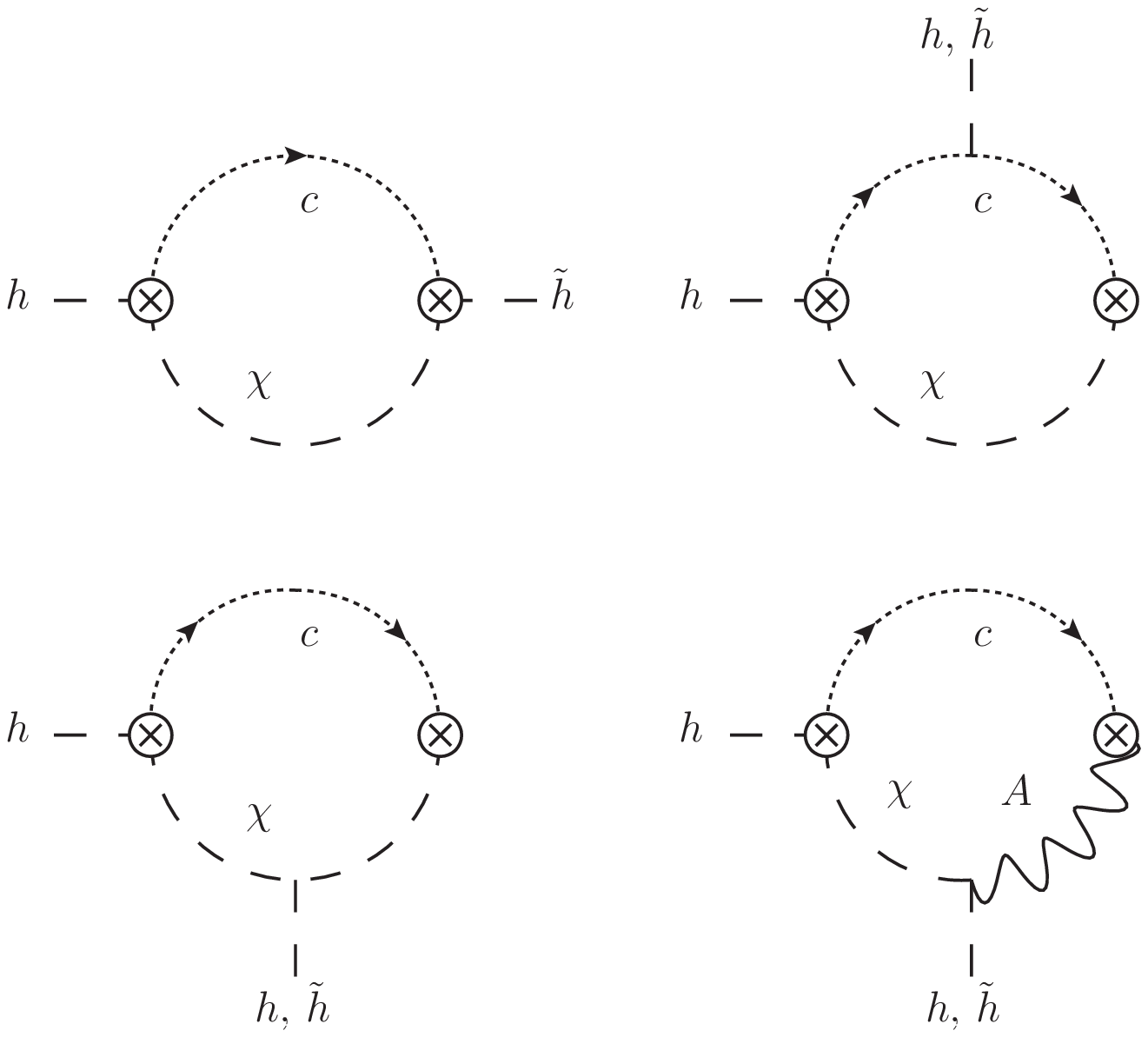}
\end{center}
\caption{
\label{fig:tD}
\small The diagrams contributing to the function $\tilde D$. }
\end{figure}
A similar calculation can be done for the factors $D$ and $\tilde D$
in (\ref{eq:C_grad_ex}). These are obtained from calculating the
diagrams shown in Fig.~\ref{fig:D} and Fig.~\ref{fig:tD}, respectively, expanding in powers of the
external momentum $p_\mu$, and extracting the contribution
proportional to $p^2$. This yields the result 
\bea\label{eq:Dterm}
D & = & \frac{g^2T\xi}{192\pi\phi}\Bigg( \frac{1}{m_\chi^3} + \frac{3m_{FP}^2}{m_\chi^5}
+ \frac{9}{m_\chi m_{FP} (m_{\chi}+m_{FP})} + \frac{8}{(m_{\chi}+m_{FP})^3} \nn\\
&& - \frac{32 m_{FP}^2}{\xi(m_\chi^2-m_{FP}^2)}\left( \frac{1}{m_{FP}(m_{FP}+m_T)^2} 
-  \frac{1}{m_{\chi}(m_{\chi}+m_T)^2} \right) \nn \\
&& + \frac{\delta}{\xi}\bigg[ \frac{4m_{FP}(m_\chi+5m_{FP})}{m_\chi(m_\chi+m_{FP})^4} 
+ \frac{6m_{FP}^2}{m_\chi^5}  \bigg] \nn\\
&& + \left(\frac{\delta}{\xi}\right)^2 \frac{m_{FP}^4(25m_\chi^3+29m_\chi^2m_{FP}+15m_\chi m_{FP}^2+3m_{FP}^3)}{m_\chi^5(m_\chi+m_{FP})^5} \Bigg) \,,\nn\\
\tilde D & = & -\frac{g^2T\xi}{96\pi}\Bigg(\frac{1}{m_\chi m_{FP} (m_{\chi}+m_{FP})}
+ \frac{1}{m_\chi^3} + \frac{4}{(m_{\chi}+m_{FP})^3}  \nn\\
&& {} + \frac{\delta}{\xi} \frac{m_{FP}^2(7m_\chi^2+4m_\chi m_{FP}+m_{FP}^2)}{m_\chi^3(m_\chi+m_{FP})^4} \Bigg) \;.
\eea

\section{Wall tension \label{sec:walltension}}

In this section we discuss the gauge dependence of the wall tension
in more detail.

First, we show that an explicit calculation reproduces the
dependence advocated in (\ref{eq:wall_gauge_dep}) by use of the
Nielsen identity. The wall tension (without including the $Z$ factor)
is given by
\be
\sigma_0 =  \int d\phi \sqrt{V_0 + \delta V} \, ,
\ee
where we split the potential into a gauge-independent piece $V_0$ and
the contribution arising from the Goldstone bosons
(\ref{eq:del_V_shift}). The contribution to the wall tension from the
broken phase is nearly gauge-independent as long as the integration
boundaries are adapted consistently. Close to the symmetric phase, two
sources of gauge dependence arises. The first stems from the upper
boundary of integration and the second from $\delta V$. Due to $\delta
V \ll V_0$ one finds
\be
\Delta \sigma_0 = \frac12 \left. \int_0^{\bar \phi} d\phi \frac{\delta V}{\sqrt{ V}} 
+ \sqrt{ V} C_0 \right|_0^{\bar \phi},
\ee
and $\bar \phi$ is arbitrary as long as $2 g^2 \xi \bar \phi^2 \gg
m_\chi^2$. Let us first evaluate this integral in the regime $\bar
\phi \ll \phi_c$. This requires $\xi \gg \lambda/2g^2$ and leads to $V
\simeq m_\chi^2 \phi^2$ and $m^2_\chi = \lambda \phi_c^2 /4 = {\rm
  const}$.  The corresponding integral can be evaluated using
\be
\int_0^{\bar x} \frac{dx}{x} \left( (1+x^2)^{3/2} - x^3 - 1 \right)
\equiv \frac32 \bar x - \log(\bar x) -\frac{4}{3} + \log2 
+ {\cal O}(\log (\bar x)/\bar x).
\ee

We are mostly interested in the two leading terms. The constant term
do not lead to a $\xi$ dependence and the subleading corrections are
of order $\lambda \log(g) /g^2 \xi \sim g \log(g) / \xi$.  In the wall
tension, one cannot chose $\bar \phi$ parametrically smaller than
$\phi_c$, but one can neglect the range of integration from zero to a
few, since it does not contribute to the two leading terms. This makes
the result meaningful, since the effective action cannot be trusted
for $x \ll 1$ due to the breakdown of the gradient expansion. This also
allows one to expand the numerator of the integrand and to include the
full mean field potential and not just its linearization in the
symmetric phase. The arising integral is
\bea 
&& \hskip -1 cm 
\int_0^{\bar x} \frac{dx}{x(1-x/x_c)} \left( \frac32 x
(1-x/x_c)(1-2x/x_c) - 1 \right) \nn \\
&\equiv& \frac32 \bar x (1 - \bar x/ x_c) - \log(\bar x) + \log (1- \bar x/x_c) \nn \\
&& \quad  + \, {\rm const } + {\cal O}(\log (\bar x)/\bar x) \, .  
\eea

After this procedure, $\bar x$ can be chosen as a fixed multiple of
$x_c$, e.g. $\bar x= x_c/2 = \sqrt{2 g^2 \xi / \lambda}$. The linear
term cancels then against the contribution from the boundary of
integration (what can be checked using (\ref{eq:C_result})) while the
logarithmic term reproduces the relation (\ref{eq:wall_gauge_dep})
\be
\Delta \sigma_0 \simeq \frac{1}{48 \pi} T_c m_\chi^2 \, \log(\xi) \, .
\ee
The constant term is gauge-independent and
corrections to this relation are of relative order $g \log(g) / \xi$.

Next, we examine if the logarithmic terms cancels against the
contributions arising from the wave function corrections $Z$. For
larger values of $x$, the integrand can be expand as
\be
\Delta \sigma_Z = \frac12 \int_0^{\bar \phi} d\phi \, \delta Z \sqrt{ V} \, . 
\ee
This integral still diverges, but one can extract a gauge-independent contribution 
\be
\label{eq:Z_limit}
Z-1 \xrightarrow{\phi \to \infty} 
\frac{g^2 T}{3 \pi} \left( \frac{11}{8 m_T} - \frac{m_T^2}{16 m_L^3} \right) \, .
\ee
We will deal with this contribution later.

Since the leading contributions from the remaining $\delta Z$ is
of order $\log(\bar x)$, the potential can be linearized from the
start and the integration of the $\xi$-dependent terms of $\delta Z$
in (\ref{eq:Z_general}) leads for large $\bar x$ to a contribution
\be
\Delta \sigma_Z \simeq -\frac{1}{12 \pi} T_c m_\chi^2 
\left( -\frac{5}{12} + \frac{4}{1-\xi} + 
 \frac{4 \log (1-\sqrt{1-\xi}) - 2 \log(\xi)  }{(1-\xi)^{3/2}} \right) \, .
\ee
Note that the right-hand side is regular for $\xi>0$, especially it has no pole
for $\xi=1$. It is also possible to estimate the $\xi$ dependence by evaluating the
$\xi$-derivative similarly to Eq.~(\ref{eq:N_wall_tension}), but including
$Z$ and using the Nielsen identity Eq.~(\ref{eq:ZC_relation}). This leads to
the same result as shown above.
For $\xi$ not too small, the $\xi$ dependence of $\Delta \sigma_0$ and
$\Delta \sigma_Z$ are of the same order and have opposite signs.
However, there is no systematic cancellation between the
$\xi$ dependences of these two quantities.  In any case, it turns out
that it is not justified to neglect the integration in $x$ from zero
to a few, since the corresponding contribution is potentially as large
as the one we just presented. This contribution cannot reliably
determined due to the breakdown of the gradient expansion close to the
symmetric phase. Ultimately this prevents us from obtaining a
gauge-independent result for the wall tension.

Finally, consider the gauge-independent piece
(\ref{eq:Z_limit}). The corresponding contribution to the wall
tension can be evaluated using
\be
\int_0^1 \left( \sqrt{1 - \frac{\alpha x^2}{(x^2 + \beta)^{3/2}}} -1 \right) 
x (1-x) \simeq 
\left\{ \begin{matrix} 
\frac14 \,\alpha  \,\, {\rm for} \, \beta=0 \\
0.0146 \, \alpha \,\, {\rm for} \, \beta\simeq 1 \\
\end{matrix} \right. \, .
\ee
Hence, the longitudinal gauge bosons contribute via the wave
function correction a term
\be
\Delta \sigma \simeq -\frac{11\sqrt2}{192\pi} g m_\chi T_c \phi_c \, , 
\ee
to the wall tension. Notice that this is of order $g^{5/2}$ and
parametrically larger than the gauge-dependent contributions we
estimated before.

\section{Integrals \label{sec:ints}}

In the following we list the used one-loop integrals.

\bea
I_B(y) &=& \frac{1}{2 \pi^2} \sum_n \int dx \, x^2 \, 
\log{(4\pi^2 n^2 + x^2 + y)} \nn \\
&\simeq& \textrm{const}\,  + \, \frac1{12} y - \frac{1}{6\pi} y^{3/2} 
+ \, {\cal O}(y^2 \log y) \, , 
\eea

\bea
J_B(y) &=& \frac{1}{2\pi^2} \sum_n \int dx x^2 \frac{1}{4 \pi^2 n^2 + x^2 + y} \nn \\
  &=& \frac{d}{dy} I_B(y) \simeq - \frac{1}{4\pi} y^{1/2} + \, {\cal O}( y \log y) \, ,  
\eea
\bea
\bar K_B(y_1,y_2) &=& \frac{1}{2\pi^2} \int dx \, x^2 \, 
\frac{1}{ x^2 + y_1} \frac{1}{ x^2 + y_2}\nn \\
&\simeq& \frac{1}{4 \pi} \frac{1}{\sqrt{y_1} + \sqrt{y_2}} + \, {\cal O}(\log y)  \, ,
\eea
\bea
\bar L_B(y_1, y_2, y_3) &=& \frac{1}{2\pi^2} \int dx \, x^2 \, 
\frac{1}{ x^2 + y_1} \frac{1}{ x^2 + y_2} \frac{1}{ x^2 + y_3} \nn \\
&\simeq& \frac{1}{4 \pi}\frac{1}{(\sqrt{y_1} + \sqrt{y_2})
(\sqrt{y_1} + \sqrt{y_3})(\sqrt{y_3} + \sqrt{y_2})} \nn \\ && 
+ \, {\cal O}(y^{-1}) \, , \nn \\ \nn \\
L_B (y) &=& \bar L_B(y,y,y) \, ,\nn  
\eea
\bea
M_B(y) &=& \frac{1}{2\pi^2} \int dx \, x^2 \, 
\frac{1}{(x^2 + y_1)^4} \simeq \frac{1}{64 \pi} y^{5/2} + {\cal O}(y^{-2}) \, .
\eea

\end{appendix}

\end{document}